\documentclass[aip,pop,amsmath,amssymb, reprint,showkeys,floatfix]{revtex4-1}
\usepackage{natbib}
\usepackage[T1]{fontenc}
\usepackage[english]{babel}
\usepackage[colorlinks=true,allcolors=blue]{hyperref}
\usepackage[pdftex]{graphicx}
\usepackage{amsmath}
\usepackage{txfonts}
\usepackage{comment}
\usepackage{setspace}
\usepackage{color}
\usepackage{verbatim}

\newcommand{\Fref}[1]{Figure~\ref{#1}}

\newcommand{\fref}[1]{Fig.~\ref{#1}}

\newcommand{\Eref}[1]{Equation~(\ref{#1})}

\newcommand{\eref}[1]{Eq.~(\ref{#1})}
\newcommand{\erefs}[2]{Eqs.~(\ref{#1})-(\ref{#2})}

\newcommand{\aref}[1]{Appendix~\ref{#1}}
\newcommand{\Mean}[1]{\overline{\bf{#1}}}
\begin{document}

\title{Turbulent Plasmoid Reconnection}
\author{F.~Widmer}
\email{widmer@mps.mpg.de}
\author{J.~B\"uchner}
\affiliation{Max Planck Institute for Solar System Research, G\"ottingen, Germany}
\affiliation{Georg-August-Universit\"at G\"ottingen, Germany}
\author{N.~Yokoi}
\affiliation{Institute of Industrial Science, University of Tokyo}

\begin{abstract}

In weakly dissipative plasmas the plasmoid instability may lead, in principle, to fast magnetic reconnection through long current sheets
(CS). On the other hand it is well known that weakly dissipative, large-Reynolds-number plasmas easily become turbulent. We address the question whether turbulence enhances the energy conversion rate of plasmoid-unstable current sheets.
For this sake we carry out appropriate numerical MHD simulations. Unfortunately, it is technically impossible to simultaneously resolve,
even on most advanced computers the relevant large-scale (mean-) fields and at the same time the corresponding
small-scale, turbulent, quantities by means of direct numerical simulations. Hence we investigate the influence of small scale turbulence
on large scale MHD processes by utilizing a subgrid-scale (SGS) turbulence model. We first verify the applicability of our SGS model. 
Then we use the SGS model to investigate the influence of turbulence on the plasmoid instability. We start the simulation with
appropriate CS equilibria of the Harris-type and force-free sheets in the presence of a finite guide field in the direction perpendicular
to the reconnection plane. We first use high resolution simulations to investigate the growth of the plasmoid instability.
Then we express the energy and cross-helicity due to the turbulence in terms of the mean fields. For this sake we obtain the mean fields
by a Gaussian filtering in the framework of a Reynolds averaging turbulence model.
This way we investigate the influence of turbulence on the reconnection rate of the plasmoid instability.
To verify the predictions of the SGS-model, the electromotive force (${\bf\cal{E}}$) is calculated for the SGS-model as well as for the coarse data obtained by filtering.
In both cases of initial CS equilibria - of Harris-type and force-free - the electromotive force points in the direction opposite to the
current flow.  
The strength of ${\bf\cal{E}}$ coincides with that obtained for the mean fields. The symmetry breakage with respect to the guide field
direction causes, however, a turbulent helicity which reduces the influence of the apparent turbulent resistivity. 
This results in a reduction of the reconnection rate of guide field plasmoid reconnection which, therefore,
is attributed to a balancing between the different physical effects related to turbulence.

\end{abstract}

\maketitle

\section{Introduction}
\label{intro}
The dynamics of the solar corona, heliosphere and astrophysics is heavily influenced by turbulence of collisionless plasmas. This is true, in particular, for the
reconnection process which converts magnetic energy into particle and plasma kinetic and thermal energy, reshaping structures such as coronal loops. Reconnection can also trigger events above active regions out of critically stressed magnetized structures. Unfortunately, the rate produced by laminar reconnection for usual collisionless space plasmas described by the Sweet-Parker model is not fast enough to explain the dynamic timescale, for example, of a solar flare.\cite{RevModPhys.82.603}
For the high Reynolds numbers, i.e. typically for the weakly collisional plasma of the solar
corona, fast magnetic reconnection could be, in principle, reached by a plasmoid instability of long current sheets.\cite{Loureiro:2007gv}
The Lundquist number $S=LV_A/\eta$ (Reynolds number for $V=V_A$) provides an approximated threshold $S_{crit}\sim 10^4$ below which a Harris-type current sheet is Sweet-Parker stable.\cite{2012APS..DPPCP8059L} 
It has been numerically confirmed that above $S_{crit}$, a plasmoid instability is triggered which leads to fast
reconnection.\cite{bhattacharjee2009fast} In most astrophysical plasmas, a guide magnetic field, which is perpendicular or oblique to the reconnection field, is observed. In solar plasmas, the plasma-$\beta$ is
small due to large guide magnetic fields. These conditions requires to investigate guide field effects for both simulations and theoretical models.\cite{2010arXiv1001.1708S} Two dimensional PIC-code
simulations of force-free CS with guide magnetic field revealed a reduction of the reconnection rate proportional to the guide field strength.\cite{Pato2015,Ricci:2003yc} In MHD simulations, finite guide magnetic
fields were shown to slow the reconnection rate. In addition, guide magnetic field effects seem
to reduce the maximum value of the reconnection rate.\cite{0004-637X-799-1-79,2015arXiv151104347W} Such reduction of the reconnection rate
was also observed in laboratory experiments.\cite{RevModPhys.82.603} A better understanding of the role of the guide
magnetic field in plasmoid reconnection is, therefore, necessary as well. \\
\indent Plasmoid instability is cascading the magnetic islands size down to small scales causing a repeated break-up of the current sheet. This highly increases the current density and the vorticity around the 'X'-lines which enhances
the reconnection rate.
Note that plasmoid instability is triggered independently on the presence of a guide magnetic field. It is
well known that high magnetic Reynolds number plasmas are prone to turbulence.\cite{Matthaeus_Lamkin_1985} The plasmoid instability might be enhanced by turbulence as
well. In this context, we investigate the influence of turbulence on the reconnection rate through plasmoid unstable long current sheets. \\
\indent Unfortunately, simulations with finite grid spacing does not allow to follow turbulence down to small scales. We investigate, therefore, the influence of turbulence on the plasmoid reconnection with a subgrid-scale (SGS) turbulence model. 
In particular, we consider a Gaussian filter formulation extended from a Reynolds-averaged Navier-Stokes turbulence model.\cite{Yokoi4}
The model reveals turbulence-related contributions to the electromotive force proportional to the mean magnetic field, the current density and the vorticity.
Turbulence is, in this model, self-generated and -sustained due to the inhomogeneities of mean fields. Energy, cross-helicity and helicity of the turbulence are statistically determined as following the Reynolds averaging rules. \\
\indent The dynamic balance between the energy and cross-helicity of the turbulence has been shown to control the rate of magnetic reconnection in anti-parallel Harris-type current sheets (CSs).\cite{Yokoi1}
In two dimensions, the turbulent helicity is negligible if the system has no mirror-symmetry breakage. From a kinetic viewpoint, the guide field is considered to provide an anisotropy of the pressure tensor
components. Mirror-symmetry is then broken in two dimensions by an out-of-reconnection-plane finite guide magnetic field. Such a situation allows for the production of a turbulent kinetic and
magnetic helicities. This latter has been presented to act against the generation of turbulent energy.\cite{YokRub} The reduction of turbulent energy in presence of large guide magnetic field was shown to reduce the reconnection rate.\cite{2015arXiv151104347W}
In this context, the consequences of an enhanced turbulent helicity 
can provide important insights on influence of a guide field on turbulent reconnection. \\
\indent In order to investigate the influence of turbulence on reconnection, it is appropriate to carry out high resolution MHD simulations. We did this by considering the plasmoid instability in Harris-type CSs with
and without finite constant guide magnetic field as well as in force-free CSs with finite guide magnetic field. In order to compute turbulence, simulation results are coarse grained using a Gaussian filter. 
The Reynolds averaged turbulence model is extended to a Gaussian filter formulation (GFF). 
The filter width is chosen to be inside the inertial range of the energy spectrum of the total field. This enable us to compute the statistical
turbulence quantities in terms of filtered variables. The GFF of turbulence allows us to investigated the prediction of the Reynolds averaged turbulence model
on the spatial localisation of the turbulent energy by cross-helicity.\cite{Yokoi4} 
Through this formulation, the applicability of the turbulence model is tested by comparing the SGS electromotive force with its statistical description. The reconnection rate of the plasmoid unstable CS obtained
from the filtered fields can then be related to the SGS turbulence.
Finally, the turbulent helicity associated with the guide magnetic field is compared to a dynamo-like source for the total magnetic energy and its influence on resistive and turbulence effects is investigated. 

\section{Resistive MHD equations}
\label{equations}

The high resolution MHD simulations are carried out by solving the following set of resistive compressible MHD equations
\begin{eqnarray}
 \frac{\partial\rho}{\partial t}&=&-\nabla\cdot(\rho\boldsymbol{V}), \label{eq:density} \\ 
      \frac{\partial\rho \boldsymbol V}{\partial t}&=&-\nabla\cdot\left[ \rho\boldsymbol{V}\otimes\boldsymbol{V}+\frac{1}{2}(p+B^2)\boldsymbol{I} 
- \boldsymbol{B}\otimes\boldsymbol{B}\right]\nonumber\\
 & &+\chi\nabla^2 \rho\boldsymbol{V},\label{eq:momentum}\\
\frac{\partial\boldsymbol{B}}{\partial t}& =&\textcolor{white}+ \nabla\times(\boldsymbol{V}\times \boldsymbol{B})+\eta\nabla^2\boldsymbol{B},\label{eq:induction}\\
\frac{\partial h}{\partial t}&=&-\nabla\cdot(h\boldsymbol V)+\frac{\gamma_0-1}{\gamma_0 h^{\gamma_0-1}}(\eta\boldsymbol{J}^2)+\chi\nabla^2h.\label{eq:entropy}
\end{eqnarray}
The symbols $\rho$, $\boldsymbol V$,  and $\boldsymbol B$
denote the mass density, velocity, and the magnetic field, respectively. The mean
entropy $h$ is used instead of the internal energy in order to have
the equation in conservative form if no sources or sinks are present. The heat effects are neglected. It is related to the thermal pressure
by the equation of state $p=2h^{\gamma_0}$. The ratio of specific heats for adiabatic conditions $\gamma_0 = 5/3$ is used. The entropy is therefore conserved
if no turbulence, Joule or viscous heating is present. 
The current density is calculated from
Amp\`ere's law as $\mu_0\boldsymbol J = \nabla\times \boldsymbol B$.
The symbol $\boldsymbol{I}$ is
the three-dimensional identity matrix.
The set of equations (\ref{eq:density})-(\ref{eq:entropy}) uses dimensionless variable for a typical
length scale $L_0$, a normalizing mass density $\rho_0$ and a magnetic
field strength $B_0$.
The normalization of the remaining variables and parameters is given by
$p_0=B_0^2/(2\mu_0)$ for the pressure and $V_{\mathrm A}=B_0/(\sqrt{\mu_0\rho_0})$ for velocities. 
The current density is normalized by $J_0=B_0/(\mu_0L_0)$, the resistivity by $\eta_0=\mu_0 L_0 V_{\mathrm A}$ and the energy by $E_0=B_0^2L_0^2/\mu_0$.
The resistivity of the plasma is constant ($\eta=10^{-3}$).
The terms containing $\chi$ are used for stabilisation of the code. They are switched on locally if
the derivative of the associated quantity (for example $\rho$ in the momentum equation) reaches a minimum (maximum).

\section{Turbulence Model and Simulations}
\subsection{Filtering or Subgrid-Scale Modelling Approach}

We divide a field quantity $F$ into the grid-scale (GS) $\overline{F}$ and subgrid-scale (SGS) $f'$ components by a filtering as
\begin{eqnarray}
	\rho= \overline\rho +\rho', \ {\bf{V}} = \Mean{V} +\boldsymbol{v}' , \ {\bf{\Omega}} = \Mean{\Omega} +\boldsymbol{\omega}', &&\\
	{\bf{B}} = \Mean{B} +\boldsymbol{b}', \ {\bf{J}} = \Mean{J} +\boldsymbol{j}', \ {\bf{E}} = \Mean{E} +\boldsymbol{e}',&&
\end{eqnarray}
where ${\bf{\Omega}}=\nabla\times{\bf{V}}$ is the vorticity and ${\bf{E}}$ is the electric field.
The filtered, or GS fields, are considered to be the mean fields.
The GS correlation between $F$ and a second field variable $G$ is denoted by
\begin{equation}
	C_{GS} = \overline{F} \ \overline{G},
\end{equation}
while the SGS counterpart is defined by
\begin{equation}
	C_{SGS} = \overline{FG}-\overline{F}\ \overline{G}. \label{eq:Csgs}
\end{equation}
If the filtering procedure has the projection property $\overline{\overline{F}}=\overline{F}$ ($\overline{f'}=0$), the
$C_{SGS}$ recovers the usual Reynolds averaging:
\begin{eqnarray}
	C_{SGS} &=& \overline{FG}-\overline{F} \ \overline{G}= \overline{\left(\overline{F}+f'\right)\left(\overline{G}+g'\right)}\nonumber \\
		       &=&\overline{\overline{F} \ \overline{G}} + \overline{\overline{F}g'} + \overline{f'\overline{G}} + \overline{\overline{f'}\overline{g'}}-\overline{F}\ \overline{G} \\
		       &=& \overline{f'g'}.
\end{eqnarray}
The chosen filter width is such that the SGS correlation is as close as possible to a Reynolds averaging (\aref{App:fitlerWidth}).
The induction equation after filtering is given as
\begin{equation}
	\partial_t \overline{\boldsymbol{B}} = \nabla\times\left( \overline{\boldsymbol{V}}\times \overline{\boldsymbol{B}}+\cal{E}\right) +\eta\nabla^2 \overline{\boldsymbol{B}},
\label{eq:IndRANS}
\end{equation}
The additional electromotive force $\cal{E}$ arising in the induction equation due to the filtering is given by
\begin{equation}
	\cal{E}\equiv\Mean{V\times B}-\Mean{V}\times\Mean{B} \label{eq:EMF}.
\end{equation}
The electromotive force can be modelled similarly to the Reynolds formulation of \citet{Yokoi2} as
\begin{equation}
	{\cal{E}}_M= -\beta \mu_0 \Mean{J}+ \gamma\sqrt{\mu_0\rho}\Mean\Omega +\alpha\Mean{B},
\label{eq:EMFY}
\end{equation}
where $\beta$ acts as a turbulent resistivity and $\gamma$ and $\alpha$ as turbulent dynamo effects. They
are considered as scalar fields and are related to the turbulent energy $K$, cross-helicity $W$ and residual helicity $H$ by the following expressions
\begin{equation}
\beta=\tau C_\beta K, \ \ \ \gamma=\tau C_\gamma W, \ \ \ \alpha=\tau C_\alpha H.
\label{eq:ModelABG}
\end{equation}
The turbulence timescale $\tau$ is algebraically related to $K$ and its dissipation rate $\epsilon$ as $\tau=K/\epsilon$. The model constants $C_\beta$, $C_\gamma$ and $C_\alpha$ are of the order $\mathcal{O}(10^{-1})$. The turbulent energy $K$, turbulent cross-helicity $W$ and turbulent residual helicity $H$ are obtained in
the GFF by
\begin{eqnarray}
	K&=& \frac{1}{2}\left[\left(\Mean{V^2}-\Mean{V}^2\right)+\frac{\left(\Mean{B^2}-\Mean{B}^2\right)}{\mu_0\bar\rho}\right], \label{eq:MathHKW1}\\
	W &=& \left(\frac{\Mean{V\cdot B}-\Mean{V}\cdot\Mean{B}}{\sqrt{\mu_0\bar\rho}}\right),\label{eq:MathHKW}\\
	H&=& -\left(\Mean{V\cdot \Omega}-\Mean{V}\cdot\Mean{\Omega}\right) + \left(\frac{\Mean{B\cdot J}-\Mean{B}\cdot\Mean{J}}{\bar\rho}\right). 
\label{eq:MathHKW2}
\end{eqnarray}

\subsection{Simulations Framework\label{Sec:Simu}}

The simulations (DNSs) are carried out for Harris-type with and without constant out-of-reconnection-plane guide field and force-free CS with finite guide fields $b_g=2$ and 5 by solving \erefs{eq:density}{eq:entropy}. The initial set-up is
$4\times3200\times12800$ grid points for a box of $0.4\times80\times320L_0^3$ in the $x\times y\times z$ directions. A system of double current sheets with periodic boundary conditions is initialised as
\begin{eqnarray}
	{\bf{B}} &=& b_g{\bf{e}}_x+B_{0}\left(\tanh{\left(y+d\right)} - \tanh{\left(y-d\right)}-1\right)\boldsymbol{e}_z, \label{eq:BzHarris}\\
h &=& \frac{1}{2}\left(1+\beta_p-\boldsymbol{B}^2\right)^{1/\gamma_0},
\end{eqnarray}
for the Harris-type CS and as
\begin{eqnarray}
\boldsymbol{B} &=&\textcolor{white}+ B_0 \sqrt{b_g^2 +\cosh^2{\left(y+d\right)}+\cosh^2{\left(y-d\right)}}\boldsymbol{e_x}\label{eq:BxForceFree}\nonumber\\
&&+B_0\left(\tanh{\left(y+d\right)} - \tanh{\left(y-d\right)}-1\right)\boldsymbol{e}_z, \label{eq:BzHarrisFF}\\
h &=& \frac{1}{2}\left(\beta_p\right)^{1/\gamma_0}, \label{eq:hFF}
\end{eqnarray}
for the force-free current sheets. Where ${\bf{b}}_g={\bf{B}}_g/|{\bf{B}}_0|$ is the constant out-of-reconnection-plane guide magnetic field and $\beta_p$ is the plasma-$\beta$. The CS are located at $\pm y\equiv \pm d = \pm 20L_0$. The initial mass density is $\rho=1$ and the initial velocity field is ${\bf{V}}=0$ for both equilibrium.
The reconnection plane is in $y\times z$, where $y$ is directed across and $z$ along the current sheet. The out-of-reconnection-plane direction is $x$. The typical length scale, magnetic field and
mass density for normalisation are given by the current sheets halfwidth $L_0$, the asymptotic mean magnetic field $B_0$ and a mean mass density $\rho_0$.
The initial perturbation for all equilibria is given by
\begin{eqnarray}
	B_y(z) = \sum\limits_{k=1}^{128}0.01\chi_1\sin\left(2\pi k \left(\frac{z}{L_z}+\chi_2\right)\right)\label{eq:BPert}
\end{eqnarray}
where $\chi_1$ and $\chi_2$ are random numbers in the range of [0,1] and $L_z$ is normalised to $L_0$ length scale in the $z$ direction. \\ \indent We obtain turbulence terms [\erefs{eq:MathHKW1}{eq:MathHKW2}] by coarse graining the full simulation data by means of a filter. While an ensemble average over many realisations would be
computationally too expensive, a time average can be used only for stationary turbulence. We choose a Gaussian filter since it conserves its properties in the transition between real to Fourier space.
Its width is chosen such that the maximum wave number $k$ cutoff lies inside the inertial range of the energy spectrum. 
For the method to be applicable, the filter width is further chosen such that it minimizes the deviations
of the SGS to the Reynolds correlation as well as to sufficiently resolve the fluctuations (see \aref{App:fitlerWidth}). This way, fluctuating
quantities are obtained from the SGS correlation $C_{SGS}$ \eref{eq:Csgs}. 
Mean quantities as the velocity, magnetic field, mass density and entropy are obtained by averaging (filtering) the fields on the fine mesh. Turbulence are obtained from the SGS correlation \erefs{eq:MathHKW1}{eq:MathHKW2}. All calculations and figures presented in the following are for the mean variables obtained by filtering high resolution simulation results. \\
\indent Since the evolution of the current sheets is dynamically non-linear and periodic boundary conditions are used, the reconnection rate
is computed using the vector potential. At each time step of the simulation, the out-of-plane mean
electric field is calculated along the current sheet when the vector potential is minimum.
Since the original long current sheet is evolving by cascading reconnection, any shorter current sheets are formed and many reconnection sites appear with time. The reconnection rate is then obtained as
the averaged mean electric field for all reconnection sites.

\subsection{Simulations Results}
\begin{figure}[h]

	\centering

	{\includegraphics[width=0.5\linewidth,keepaspectratio]{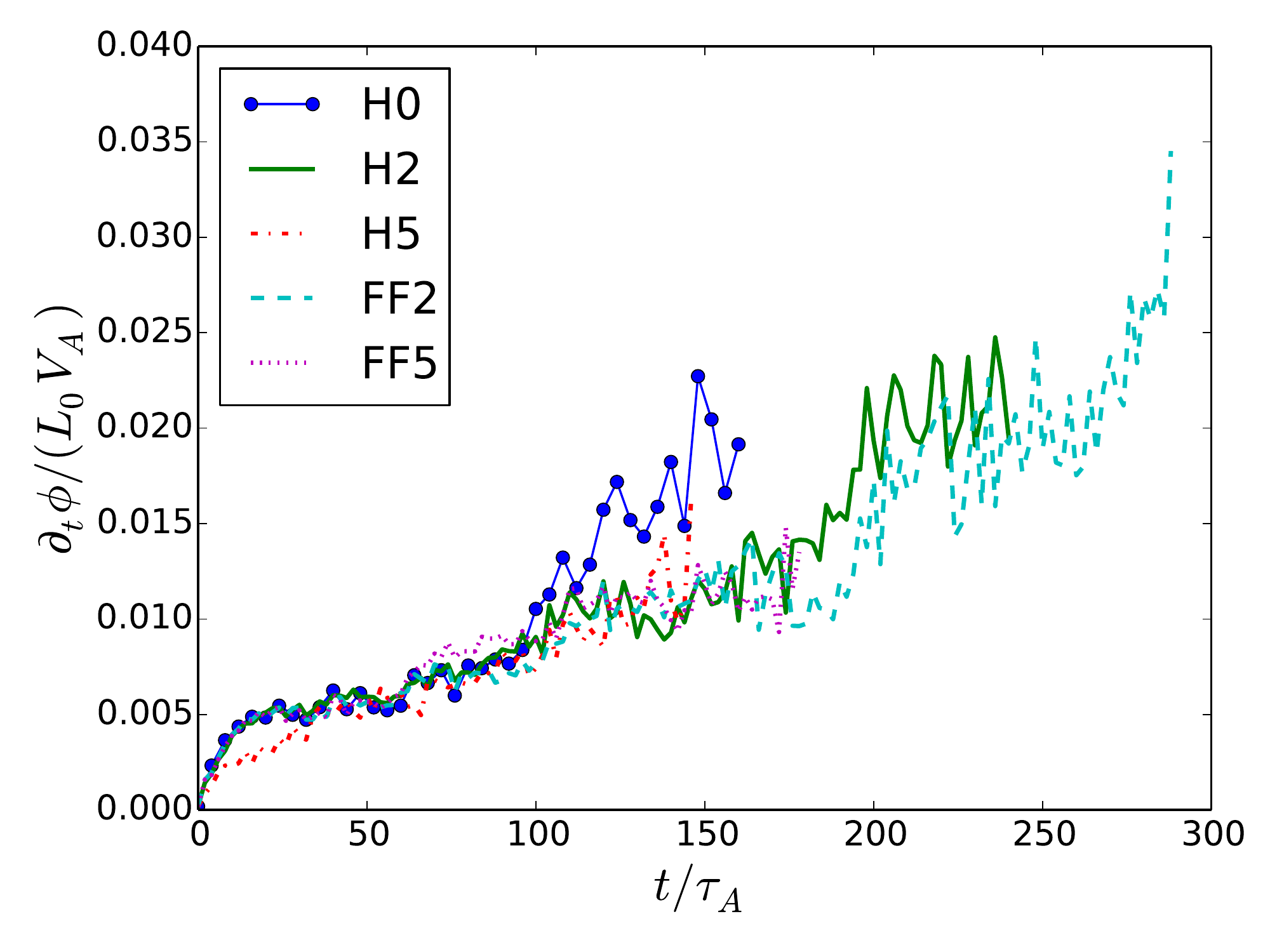}} 
	\caption{Time history of the reconnection rate in for: H0 Harris-type CS $b_g=0$, H2  Harris-type CS $b_g=2$, H5 Harris-type CS $b_g=5$, FF2 force-free CS $b_g=2$ and  FF5 force-free CS $b_g=5$.}
	\label{fig:RecRateAll}
\end{figure}
\begin{figure}[h]
  \centering
  \begin{tabular}{cc}
		\hspace{-0.5cm} {\includegraphics[width=0.5\linewidth,keepaspectratio]{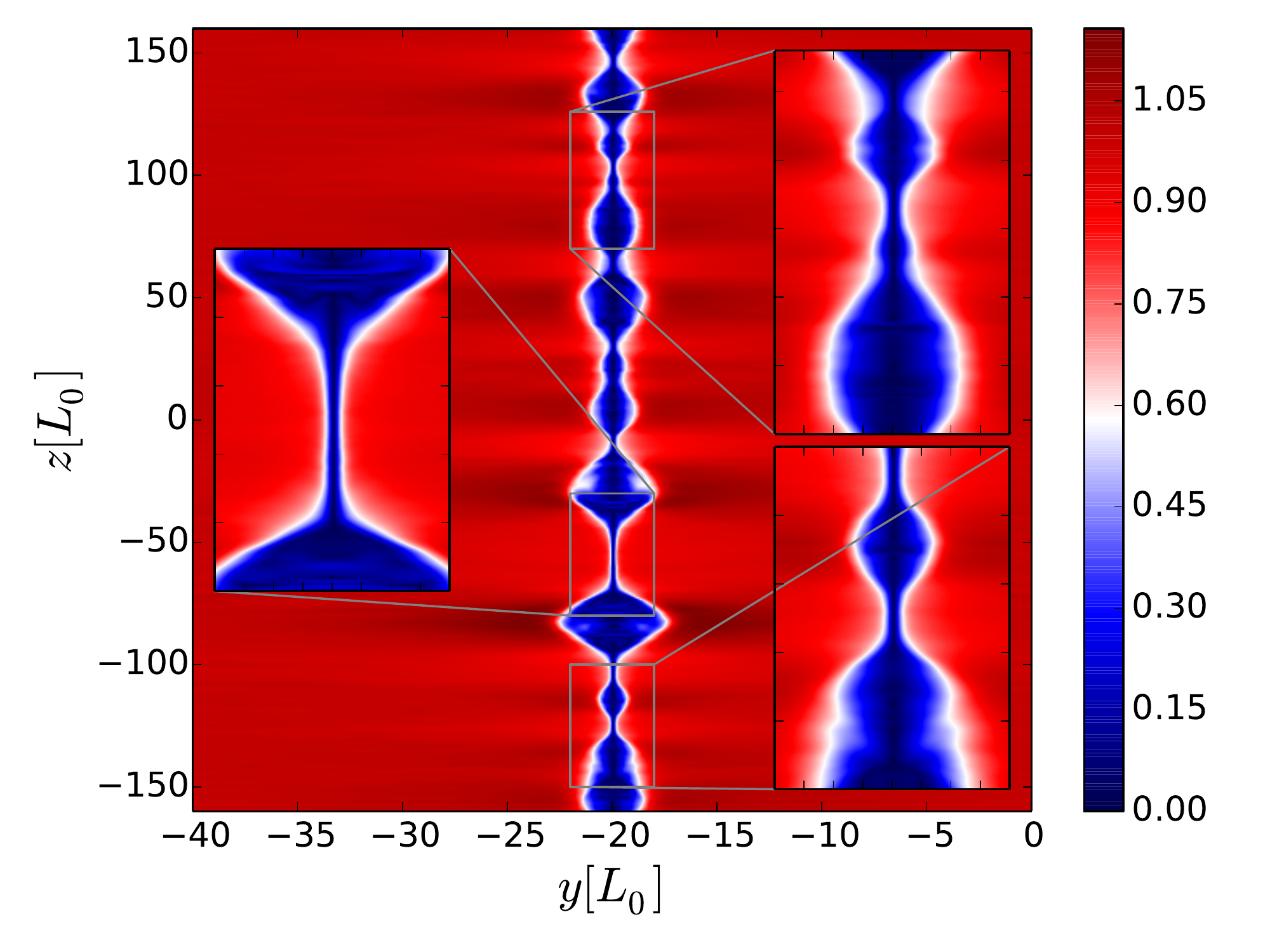}} &{\includegraphics[width=0.5\linewidth,keepaspectratio]{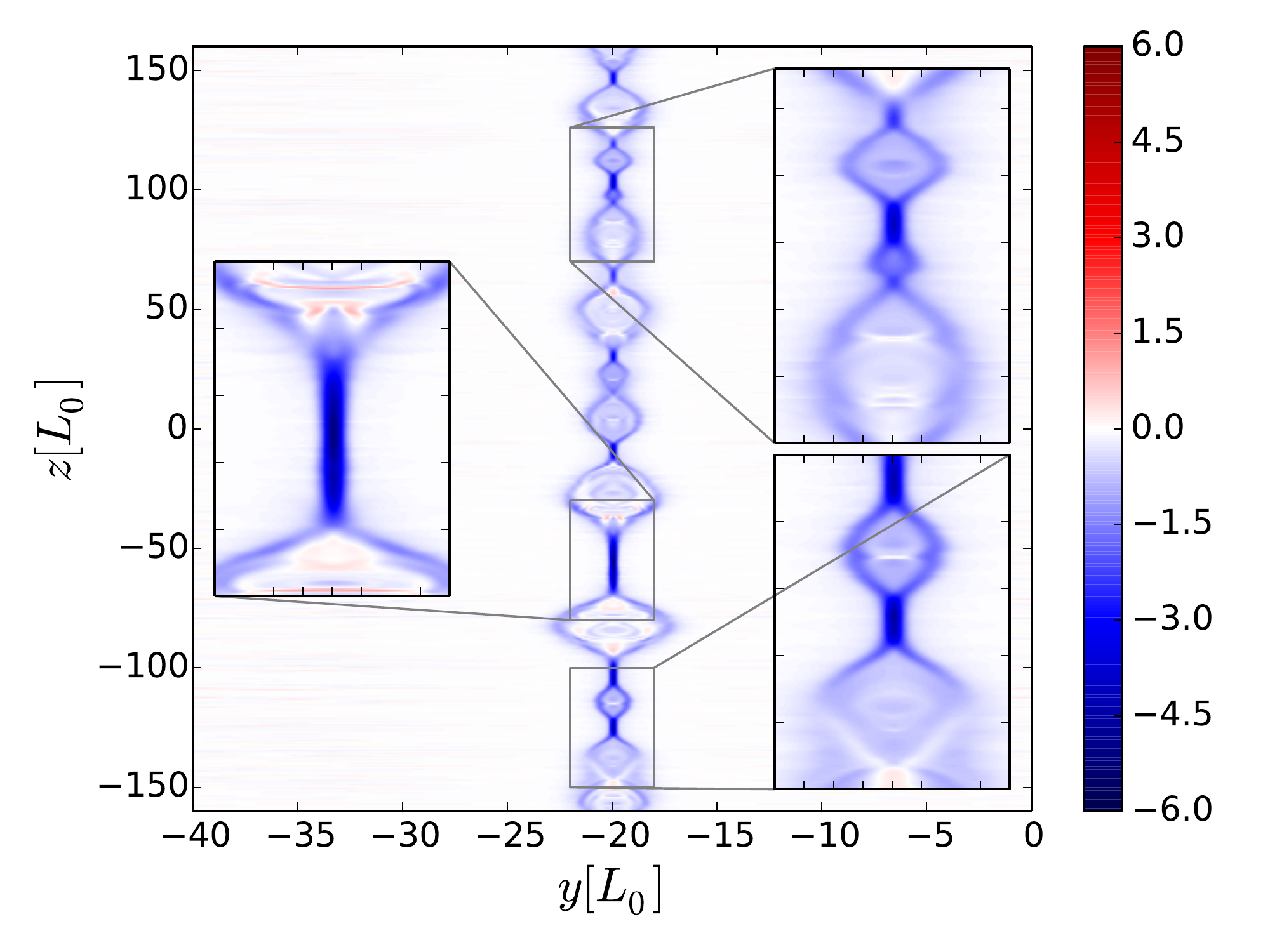}}\\
	  
	  a) Magnetic field $\bf{B}$ & b) Current density $\bf{J}$\\
		\hspace{-0.5cm} { \includegraphics[width=0.5\linewidth,keepaspectratio]{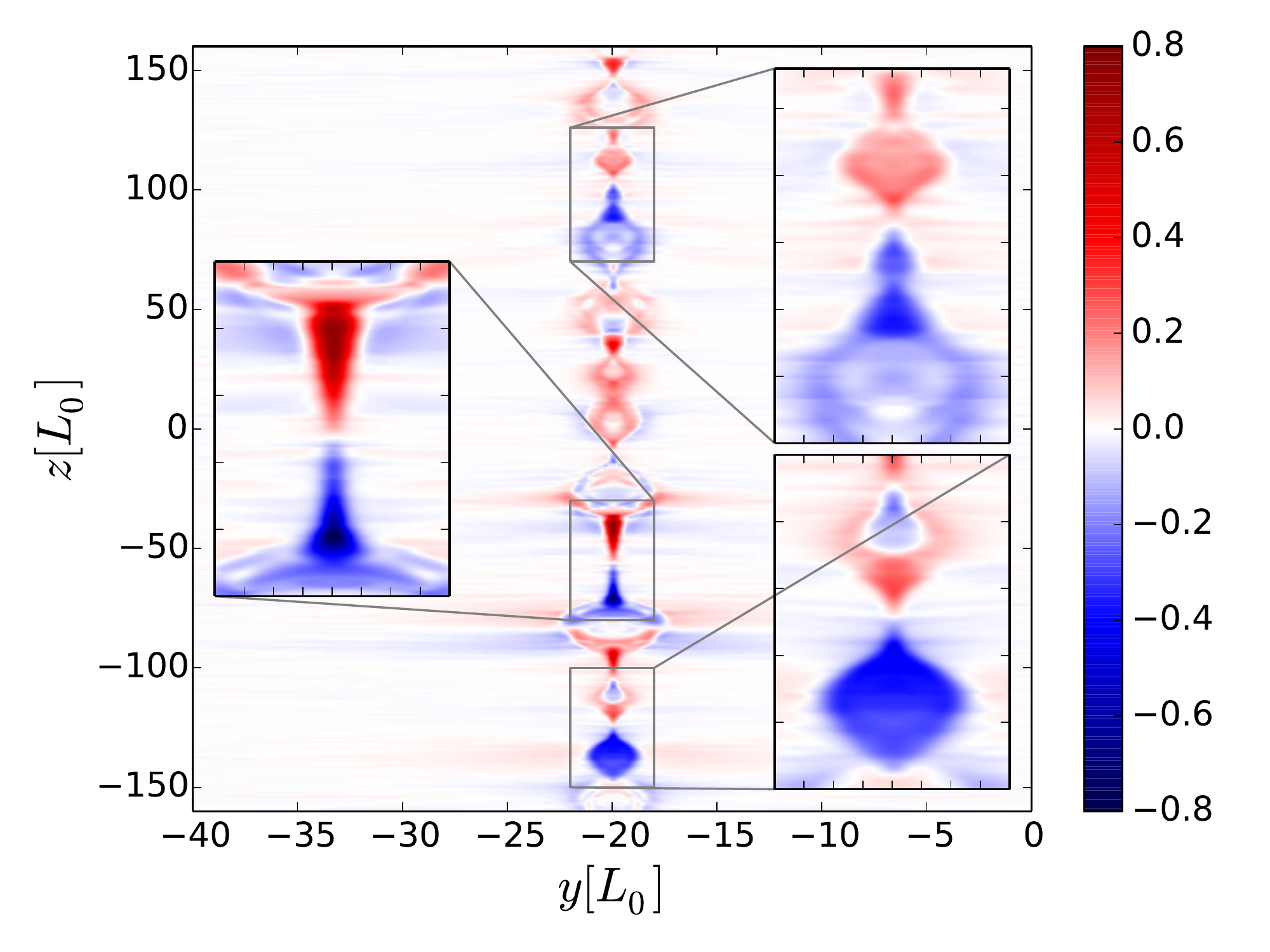}} & {\includegraphics[width=0.5\linewidth,keepaspectratio]{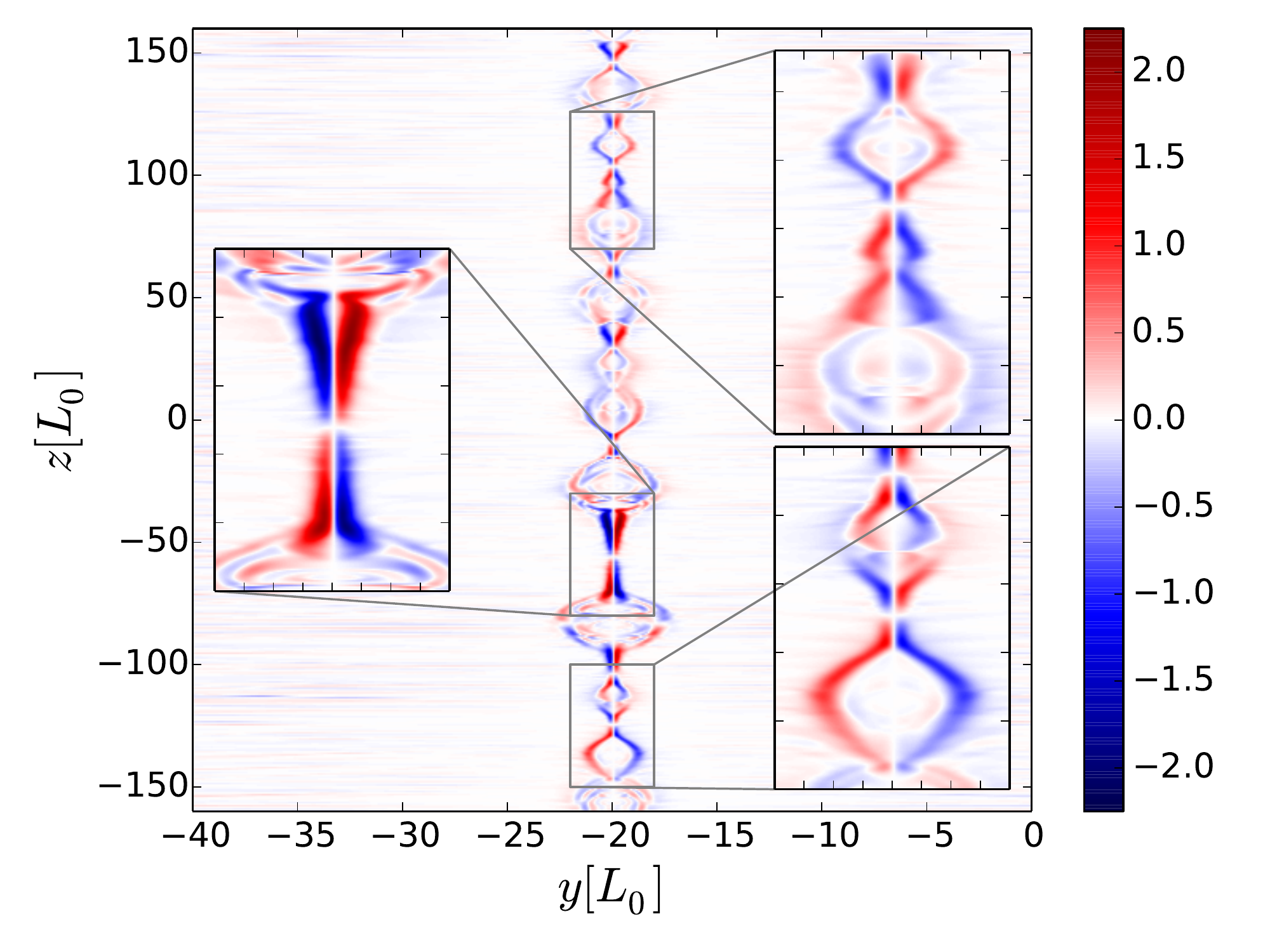}}\\
	
	  c) Outflow velocity $\bf{V}_z$ & d) Vorticity $\bf{\Omega}$\\
	\end{tabular}
\caption{Spatial distribution of the filtered field at $t=100\tau_A$ in the Harris-type CS equilibrium with $b_g=0$}
\label{fig:ContourHarris}
\end{figure}

In the first $100t/\tau_A$, the reconnection is similar in all equilibria. It then reaches a higher value sooner for a anti-parallel Harris-type CS ($b_g=0$).
The reconnection rate for in guide field reconnection finally reaches a comparable value as in Harris-type CS with $b_g=0$.
We compare the spatial localisation of the mean magnetic field magnitude (\fref{fig:ContourHarris} a) ), current density in the out-of-plane direction (\fref{fig:ContourHarris} b) ), out-flow velocities (\fref{fig:ContourHarris} c) ) and the out-of-plane vorticity (\fref{fig:ContourHarris} d) ) at $t=100\tau_A$.
The smaller reconnection rates in the first 100 $t/\tau_A$ in force-free equilibria can be related to the slightly lower maximum
value of the current density $\boldsymbol{J}$ and mean vorticity $\boldsymbol{\Omega}$. The amplitude of these mean variables
represent the stress felt by the mean magnetic and velocity fields, i.e., the strength of the gradients on these mean fields. 
\begin{figure}[h]
  \centering
  \begin{tabular}{cc}
	
		\hspace{-0.5cm} {\includegraphics[width=0.5\linewidth,keepaspectratio]{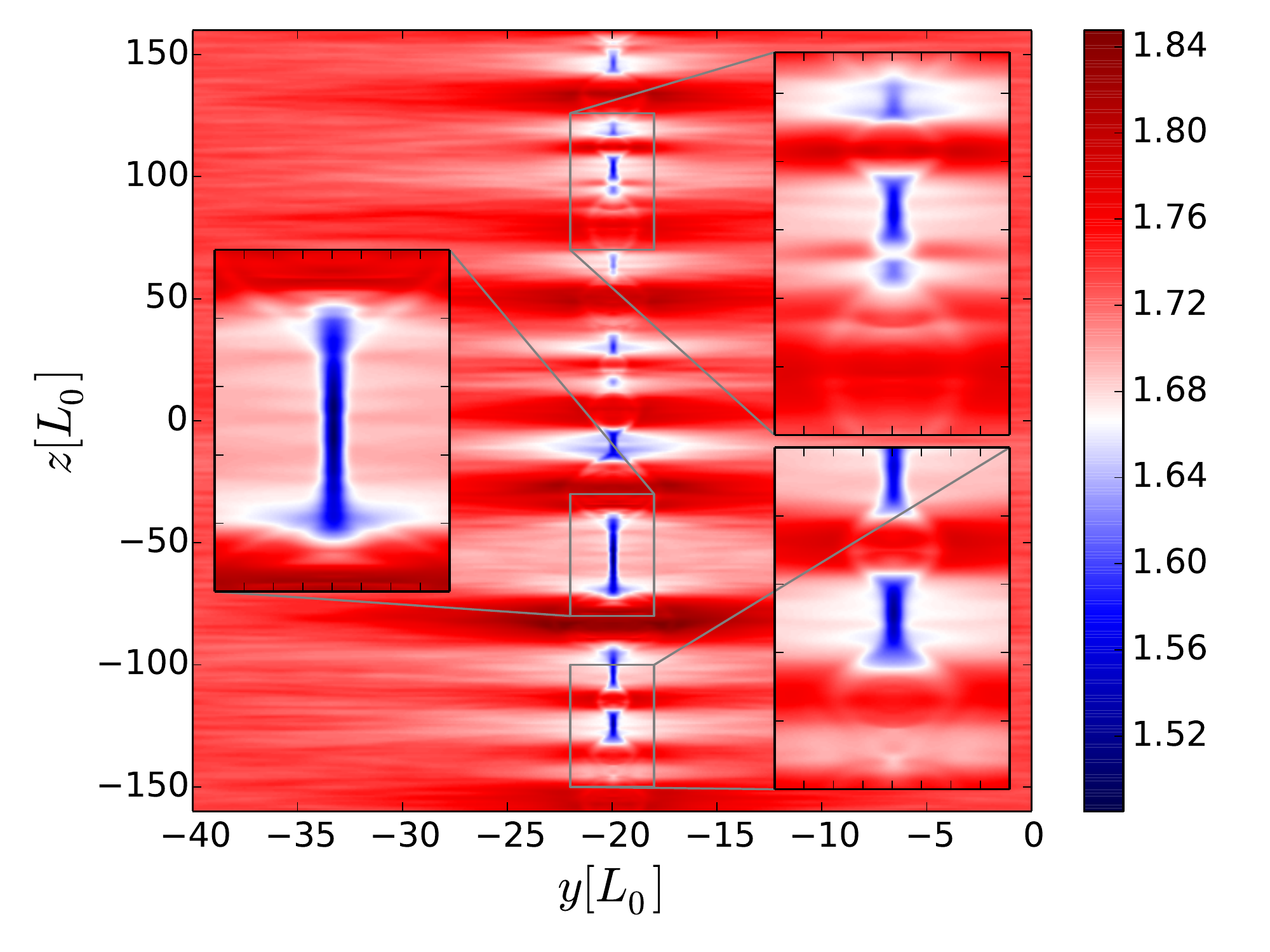}} &{\includegraphics[width=0.5\linewidth,keepaspectratio]{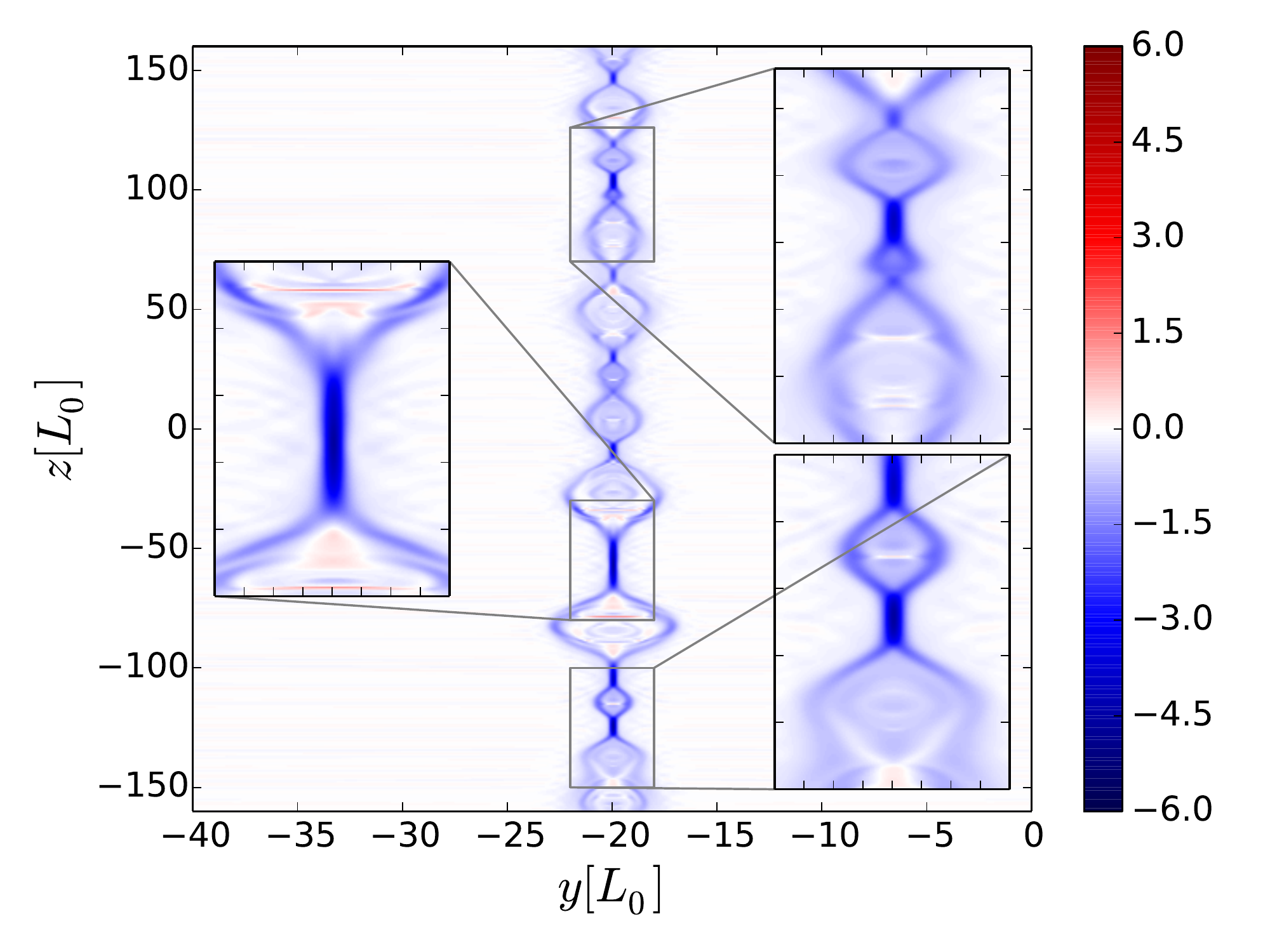}}\\
	  \hspace{-0.5cm} a) Magnetic field $\bf{B}$& b) Current densitiy $\bf{J}$\\

		\hspace{-0.5cm} {\includegraphics[width=0.5\linewidth,keepaspectratio]{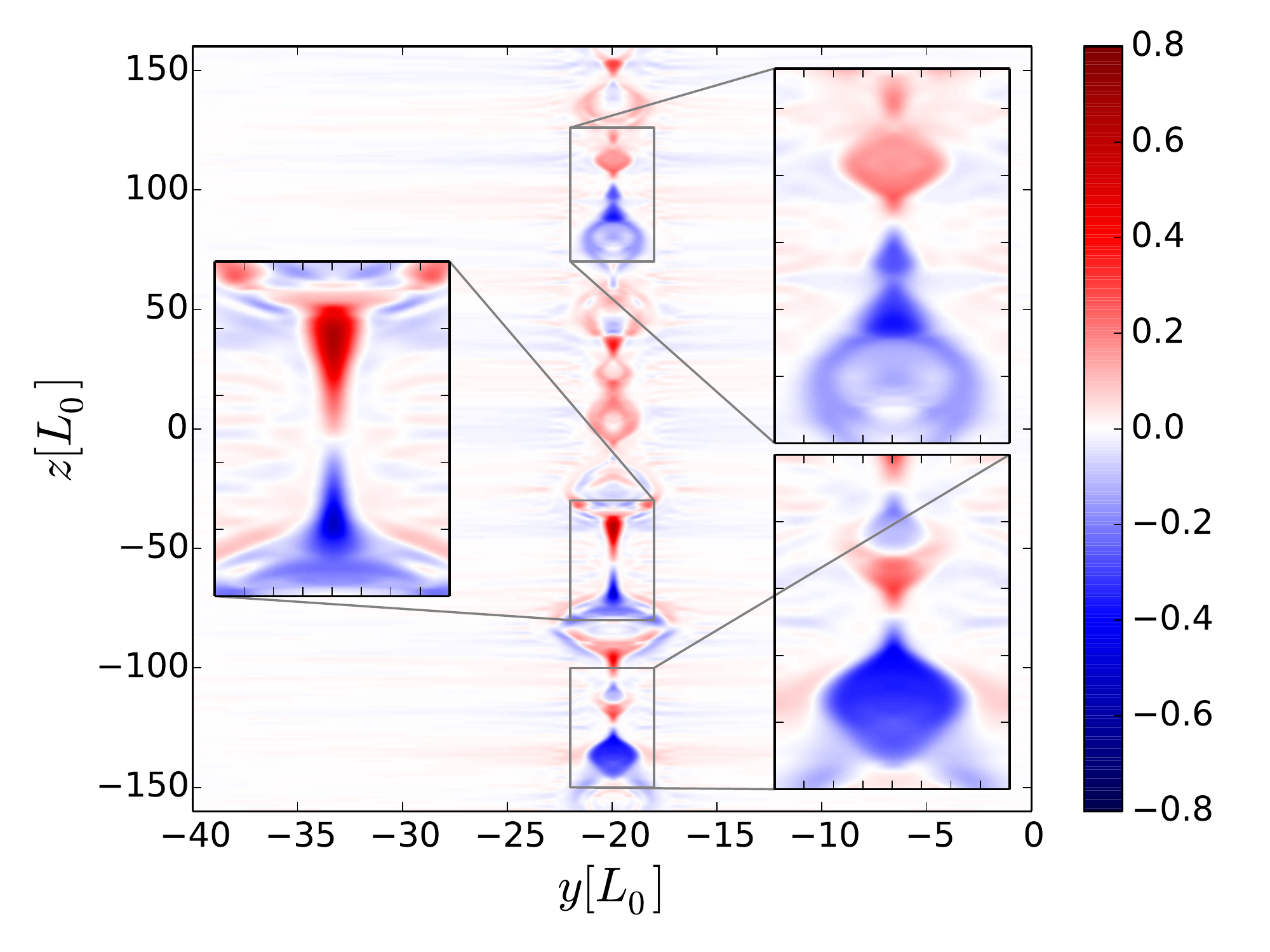}} & { \includegraphics[width=0.5\linewidth,keepaspectratio]{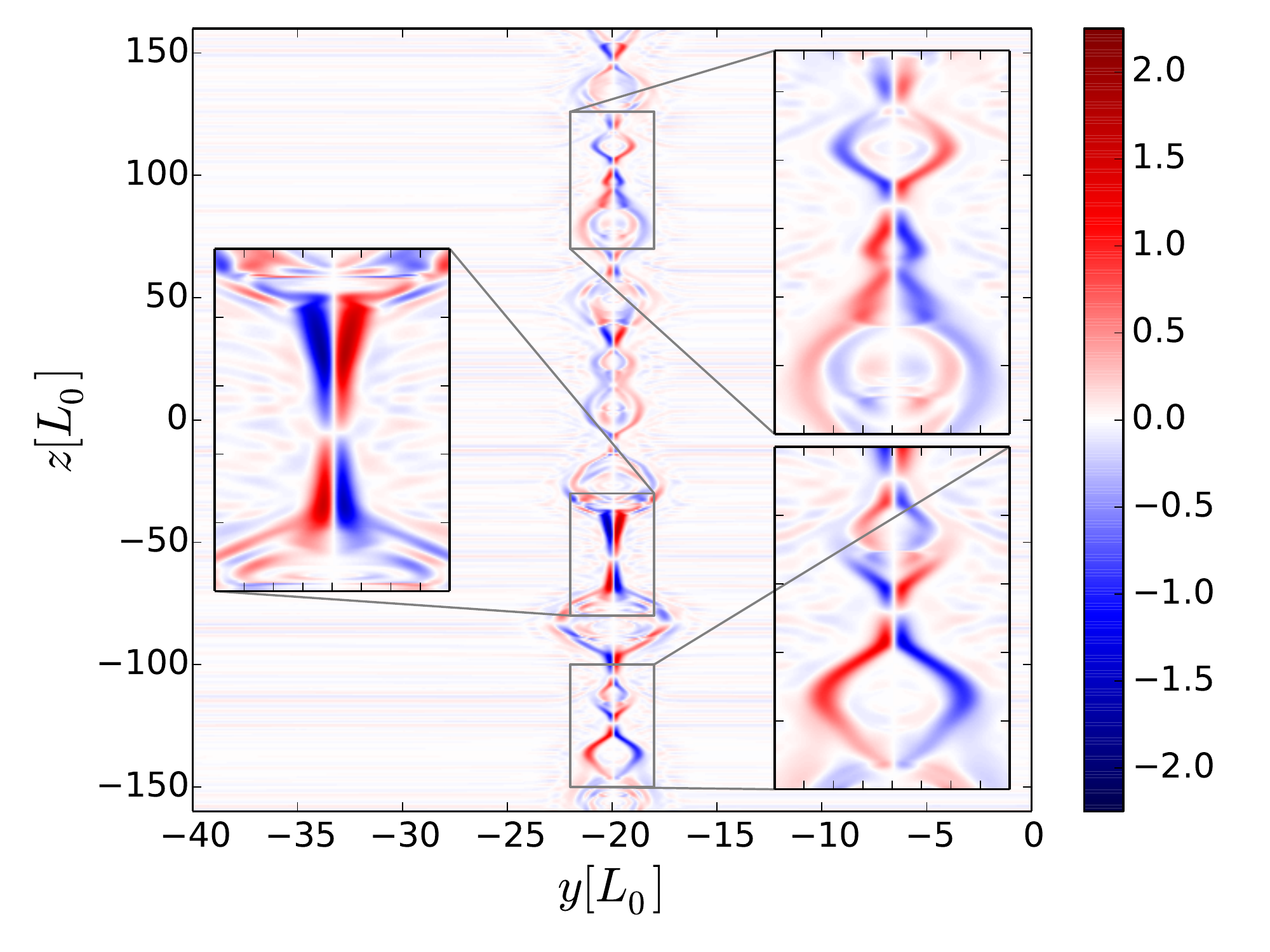}}\\
	  \hspace{-0.5cm} c) Outflow velocity $\bf{V}_z$& d) Vorticity $\bf{\Omega}$\\
	\end{tabular}
\caption{Spatial distribution of the filtered field at $t=100\tau_A$ in the force-free CS equilibrium $b_g=2$.}
\label{fig:ContourFF2140}
\end{figure}
Magnetic reconnection releases critically stressed magnetic fields and the stress strength at 'X'-point locations is related to the
reconnection rate.\cite{2007PhPl...14k2905T} A Harris-type CS equilibrium is unchanged by an additional out-of-plane constant guide magnetic field. On the other hand, a force-free current sheet has an initial in-plane current density which is reduced by
the addition of a constant out-of-plane guide field. Hence, an increase of the guide magnetic field strength reduces the Lorentz force component due to the in-plane currents but not its total amplitude.
In most astrophysical plasmas a guide magnetic field can exceed the anti-parallel reconnection magnetic field component (e.g.: in the solar corona). The reconnection rate can be estimated by dimensional analysis of the Lorentz force (\aref{App:LorForce}).
At the boundary layer of the CS where the electric field identically vanishes $\overline{\bf{J}}\approx \overline{\bf{V}}\times\overline{\bf{B}}/\eta$. The guide magnetic field
influence on the reconnection rate can be described as 
\begin{equation}
	M_{A,b_g}=M_A\left(\frac{\overline{B_z^2}}{\overline{B_x^2}+\overline{B_z^2}}\right)^2,
	\label{eq:MaBg}
\end{equation}
where $\overline{B_x}$ is the out-of-plane component of the magnetic field (guide field) and $\overline{B_z}$ the reconnecting component of the magnetic field.
$M_A$ represents the estimated value of the reconnection rate when no guide magnetic field is considered. A larger guide magnetic field decreases the reconnection
rate as found in our simulations. This was also observed in other numerical simulations and laboratory experiment.\cite{0004-637X-799-1-79,Ricci:2003yc,2013PhPl...20e5705T} \\  
\indent A turbulent helicity can be generated due to guide magnetic field effects. Hence, the guide magnetic field can be related to turbulence by the turbulent energy,
 turbulent cross-helicity and turbulent helicity.
The influence of the magnetic stress on the mean magnetic and velocity fields and the turbulent reconnection rate by turbulence is discussed in the following sections.

\section{Effect of turbulence on plasmoid reconnection}
\begin{figure}[h]
	\centering
	\begin{tabular}{cc}
		\hspace{-0.5cm} {\includegraphics[width=0.5\linewidth,keepaspectratio]{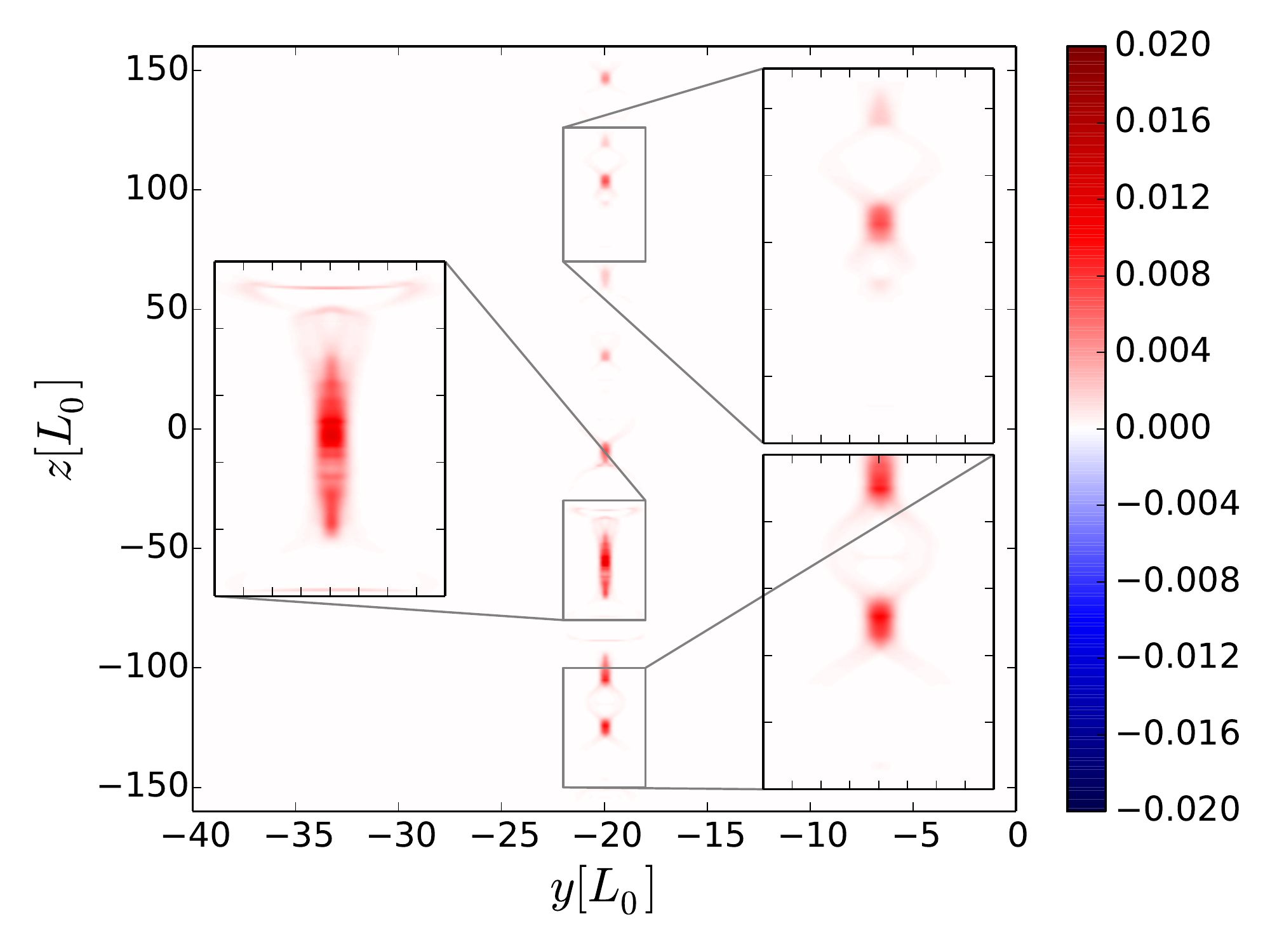}} &{\includegraphics[width=0.5\linewidth,keepaspectratio]{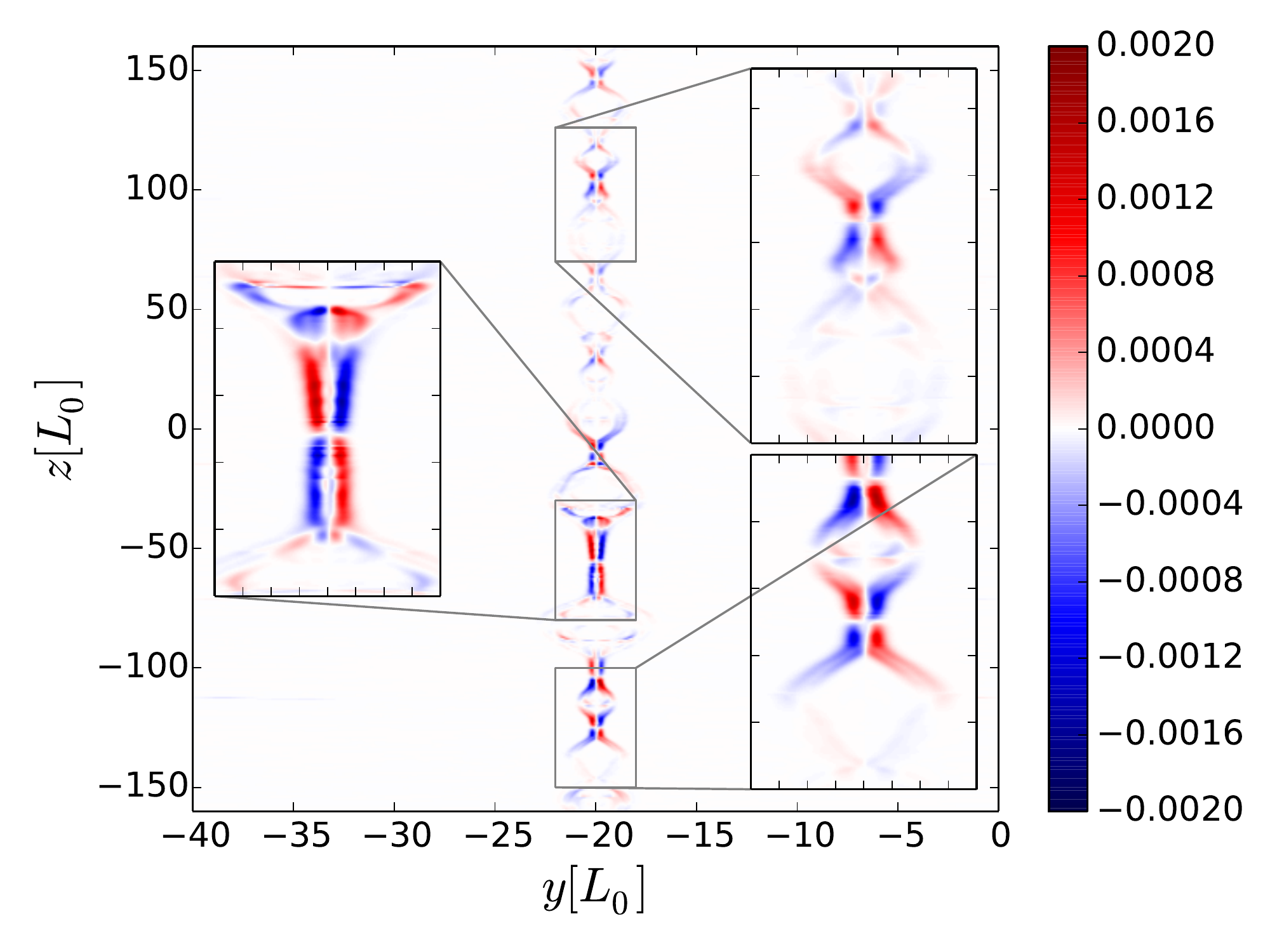}}\\
	
		\hspace{-0.5cm} a) Turbulent energy $K$ & b) Turbulent cross-helicity $W$\\
	
		\hspace{-0.5cm} {\includegraphics[width=0.5\linewidth,keepaspectratio]{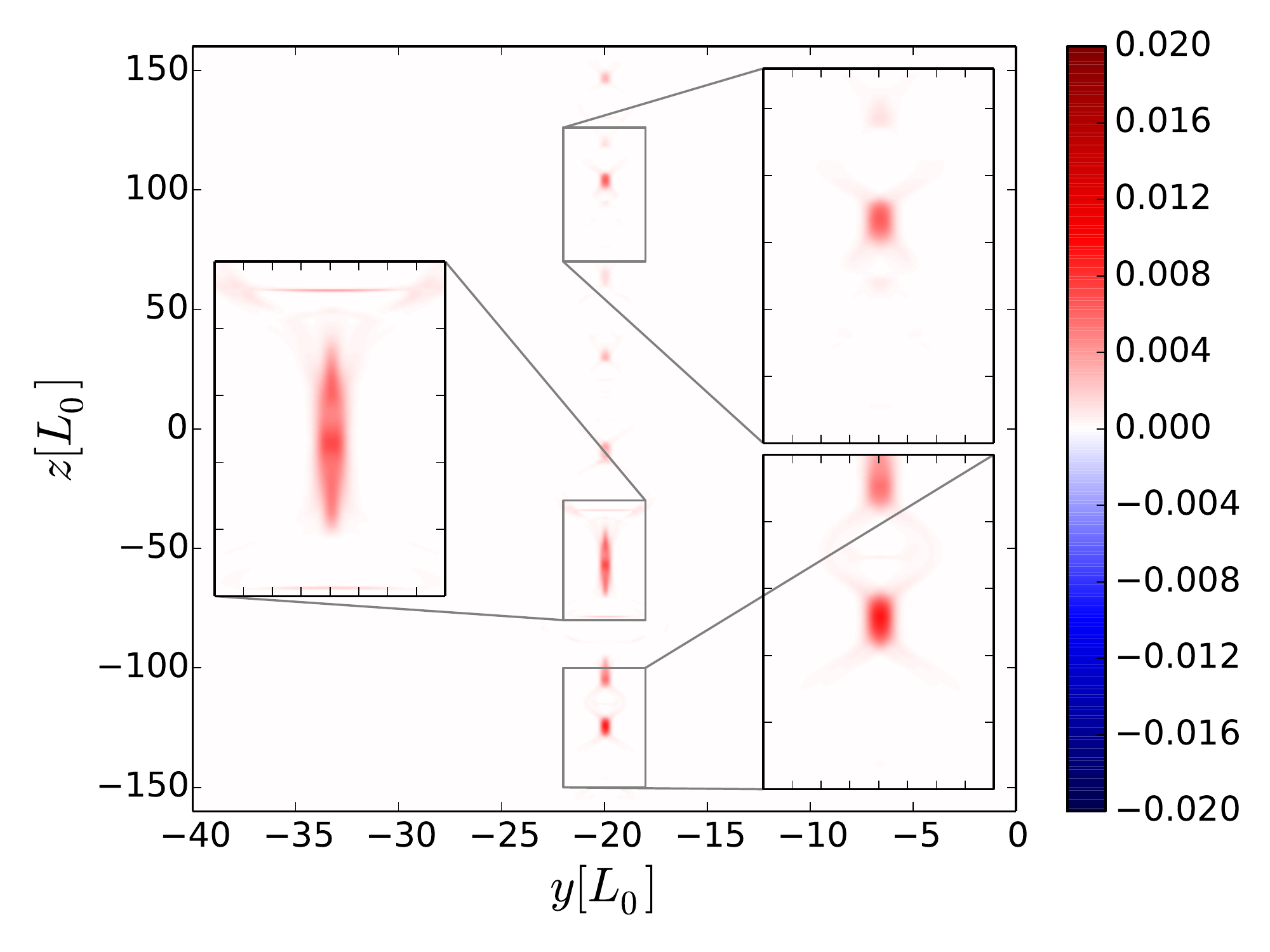}} & {\includegraphics[width=0.5\linewidth,keepaspectratio]{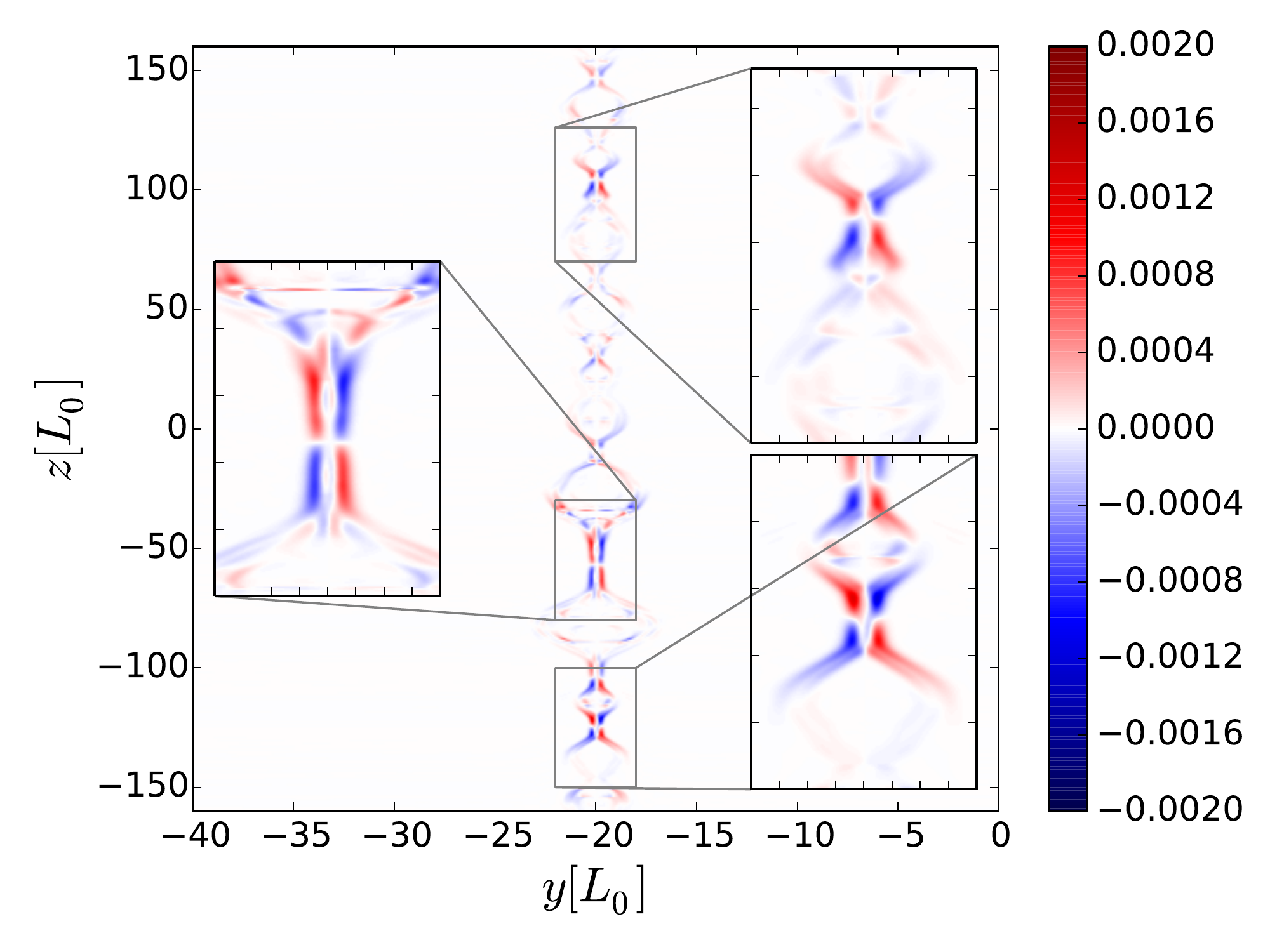}}\\
		\hspace{-0.5cm} c) Turbulent energy $K$& d) Turbulent cross-helicity $W$\\	
	\end{tabular}
	\caption{Spatial distribution of the turbulent energy $K$, cross-helicity $W$ at $t=100\tau_A$ in the Harris with $b_g=0$ ( a) and b) ) and force-free with $b_g=2$ ( c) and d)).}
\label{fig:SpatialTurbuKW}
\end{figure}
\begin{figure}[h]
	\centering
	\begin{tabular}{cc}
		\hspace{-0.5cm} {\includegraphics[width=0.5\linewidth,keepaspectratio]{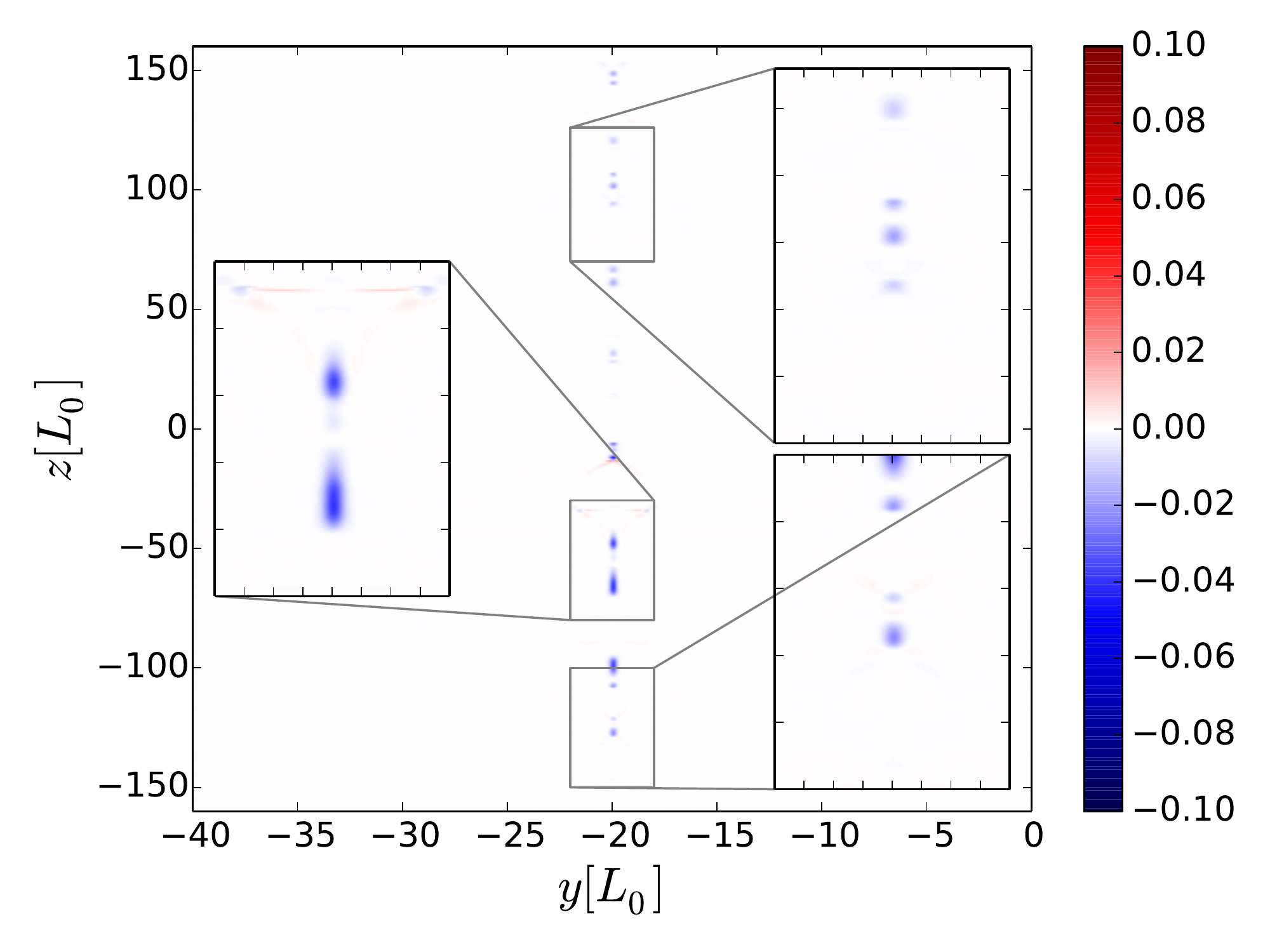}} & {\includegraphics[width=0.5\linewidth,keepaspectratio]{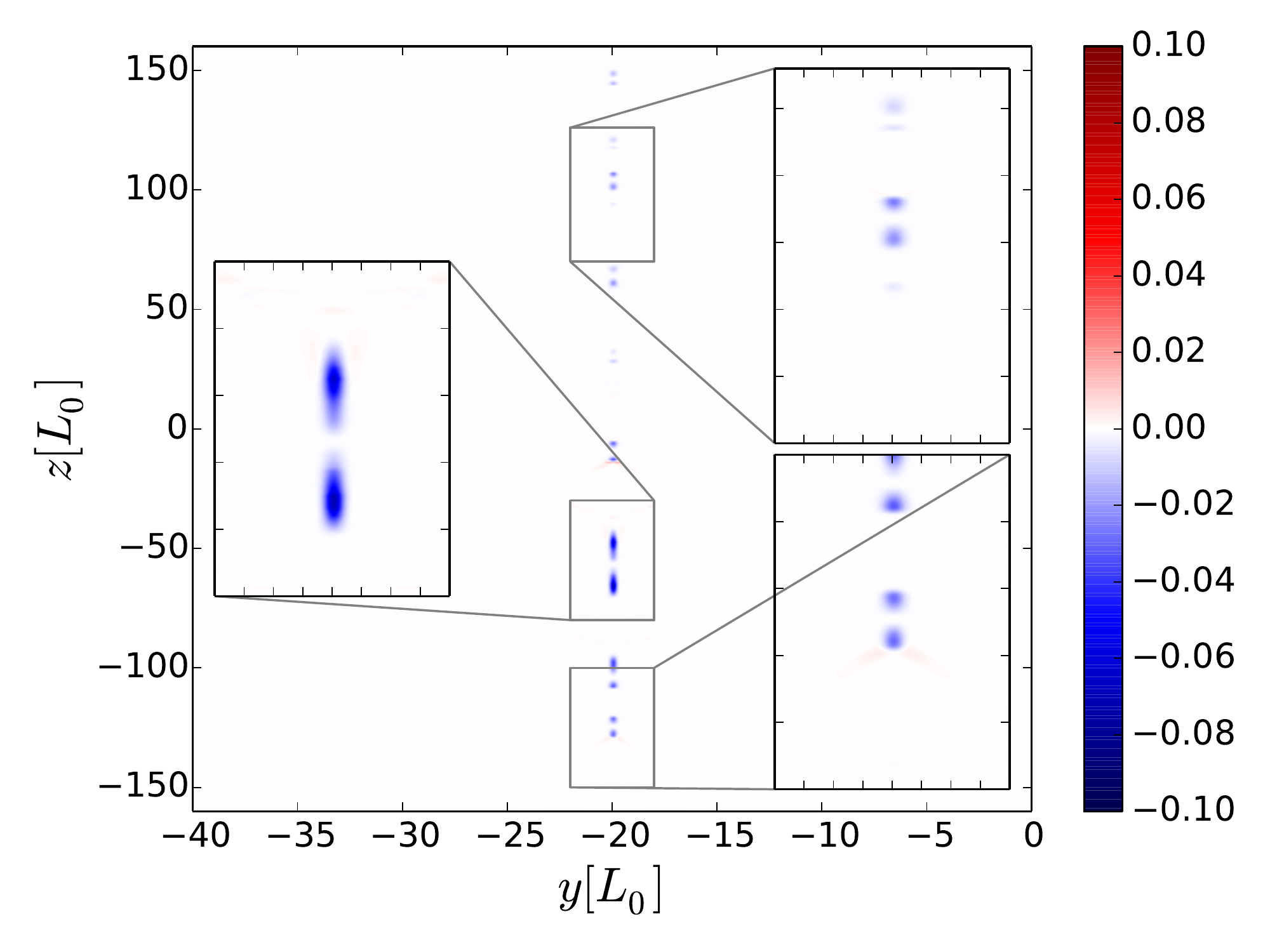}}\\
		\hspace{-0.5cm} a) $H_{kin}$ Harris $b_g=2$& b) $H_{kin}$ force-free $b_g=2$ \\
		\hspace{-0.5cm} {\includegraphics[width=0.5\linewidth,keepaspectratio]{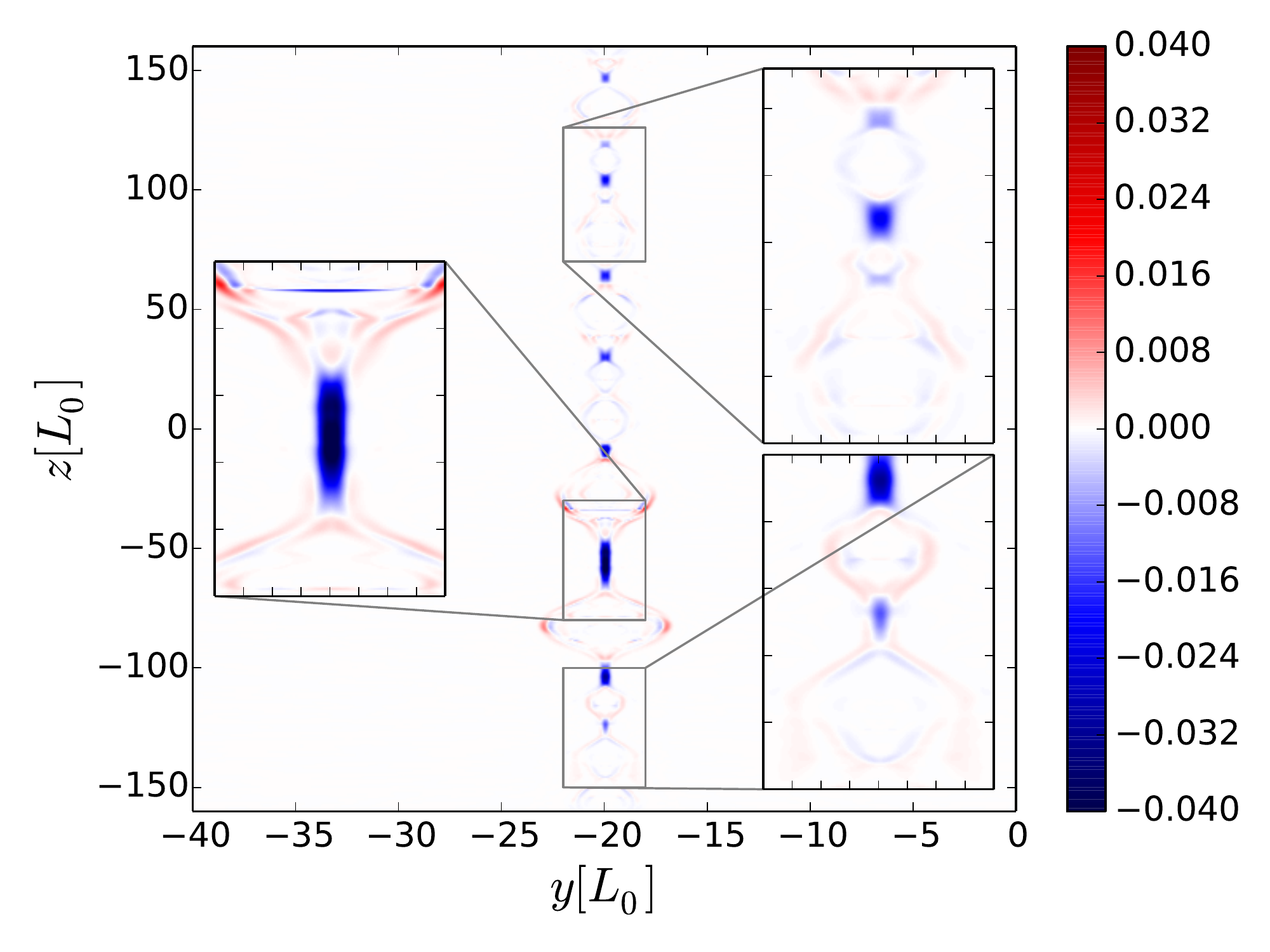}} & {\includegraphics[width=0.5\linewidth,keepaspectratio]{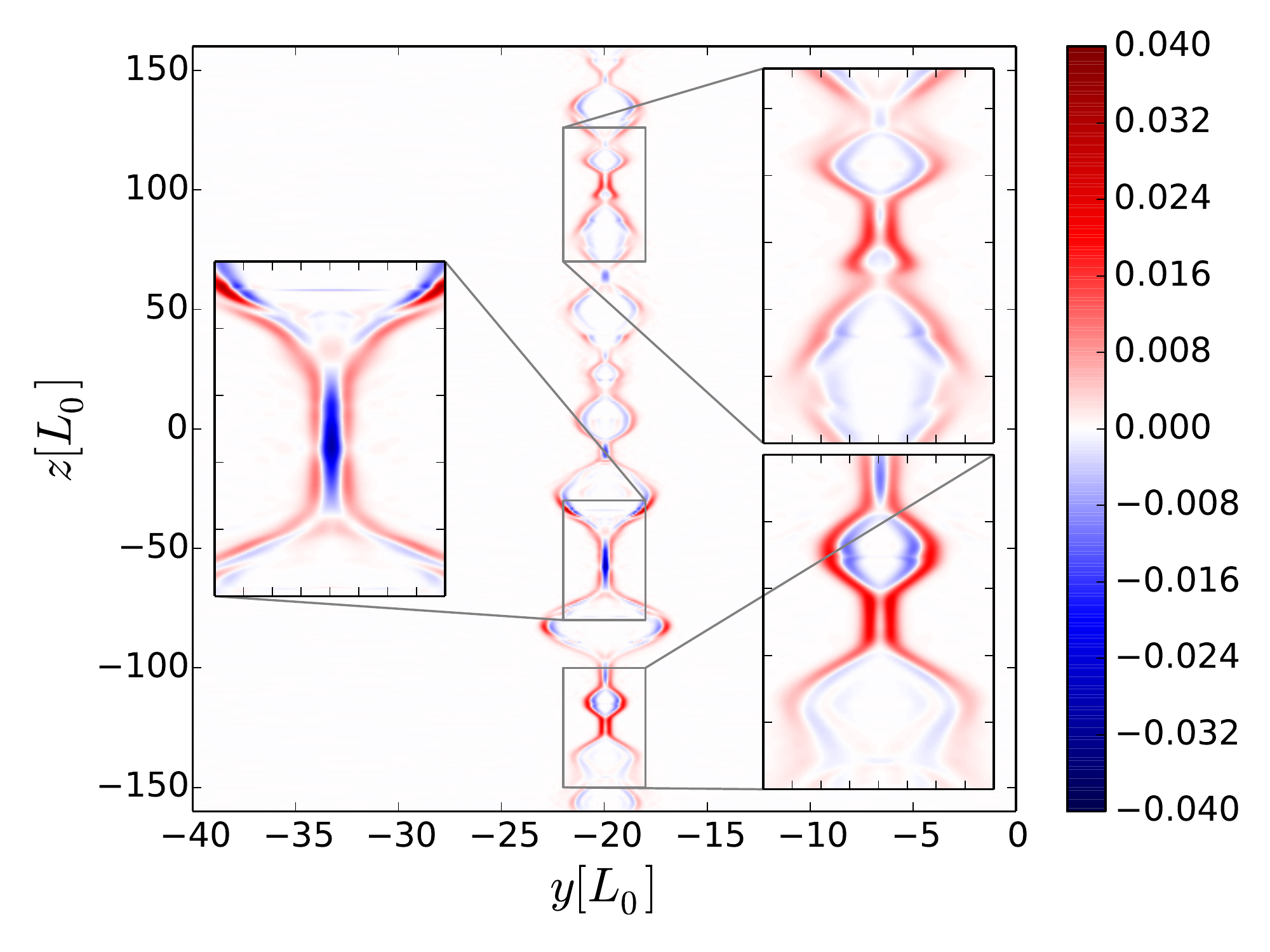}}\\
		\hspace{-0.5cm} c) $H_{mag}$  Harris $b_g=2$ & d) $H_{mag}$ force-free $b_g=2$\\
		\hspace{-0.5cm} {\includegraphics[width=0.5\linewidth,keepaspectratio]{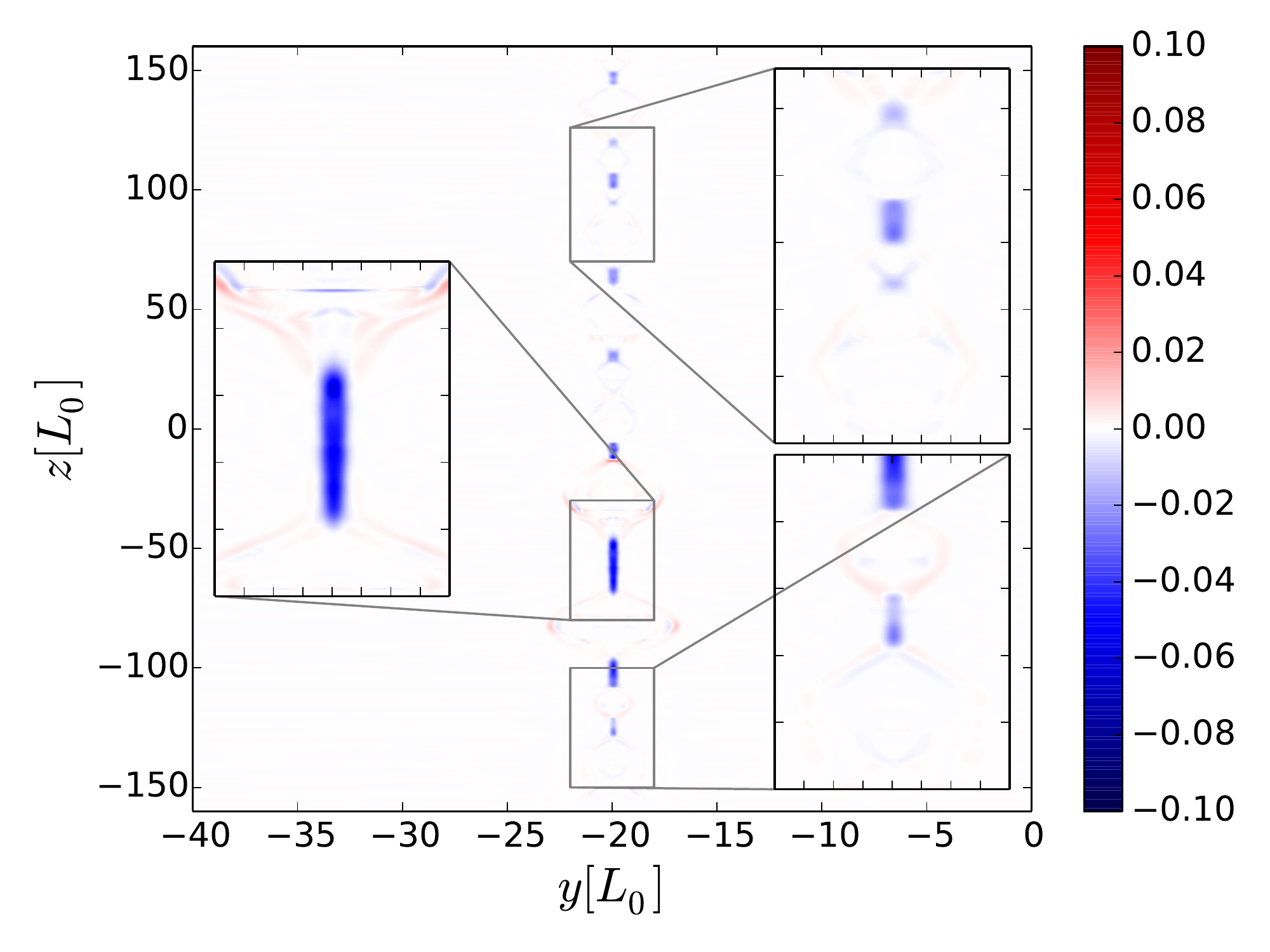}} & {\includegraphics[width=0.5\linewidth,keepaspectratio]{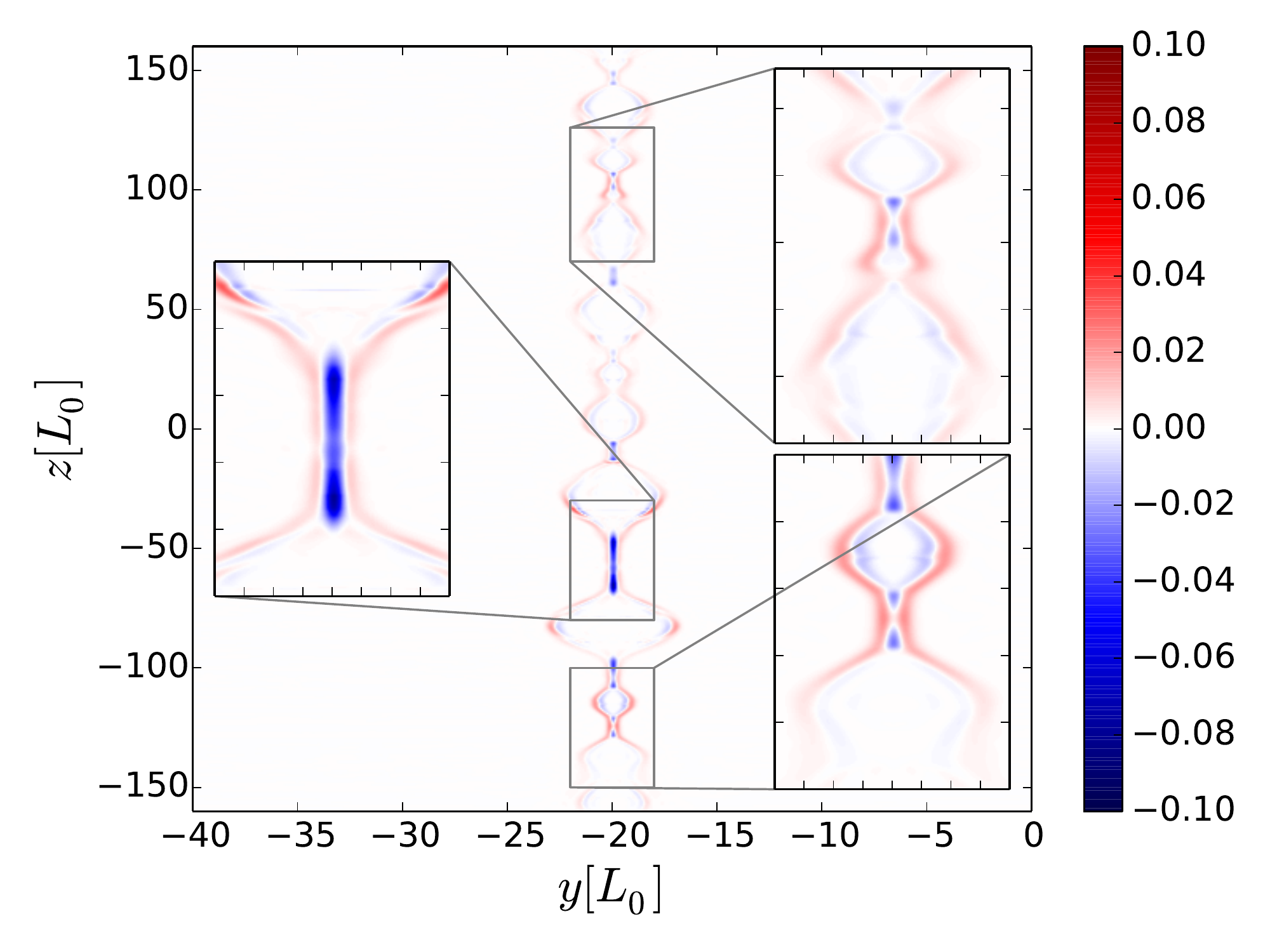}}\\
		\hspace{-0.5cm} e) $H$ Harris $b_g=2$ & f) $H$ force-free $b_g=2$\\
		\end{tabular}
		\caption{Spatial distribution of the turbulent kinetic helicity $H_{kin}$, magnetic helicity $H_{mag}$ and the total residual helicity at $t=100\tau_A$ in Harris with $b_g=2$ ( a), c) and e))  and force-free with $b_g=2$ ( b), d)  ) and f)). }
\label{fig:SpatialTurbuHel}
\end{figure}
The dynamic balance of the turbulence quantities in the course of plasmoid unstable CSs is investigated by calculating the mean turbulent
energy, turbulent cross-helicity and turbulent helicity obtained for the Gaussian filter formulation from the RANS turbulence model.
The mean electric field equation is modified by the SGS model electromotive force ${\cal{E}}_M$ 
following \eref{eq:EMFY} as
\begin{equation}
\overline{\bf{E}}= -\overline{\boldsymbol{V}}\times\overline{\boldsymbol{B}} +\left(\eta+\beta\right)\overline{\bf{J}}-\gamma\overline{\boldsymbol{\Omega}}-\alpha\overline{\boldsymbol{B}}
\label{Eq:EYok}
\end{equation}
leading to the following mean induction equation
\begin{equation}
	\partial_t \overline{\boldsymbol{B}} = \nabla\times\left(\overline{\boldsymbol{V}}\times\overline{\boldsymbol{B}}+\gamma\overline{\boldsymbol{\Omega}}+\alpha\overline{\boldsymbol{B}}-\beta\overline{\boldsymbol{J}}\right) +\eta\nabla^2\overline{\boldsymbol{B}}.
\label{eq:IndYok}
\end{equation}
\Eref{Eq:EYok} is used to obtain the current density $\overline{\bf{J}}$ which crossed with the mean magnetic field $\overline{\bf{B}}$ yields the following mean Lorentz force
\begin{equation}
	\overline{\bf{J}}\times\overline{\bf{B}}=\frac{1}{\eta+\beta}\left(\overline{\bf{E}}\times\overline{\bf{B}} +\left(\overline{\bf{V}}\times\overline{\bf{B}}\right)\times\overline{\bf{B}}+\gamma\overline{\bf{\Omega}}\times\overline{\bf{B}} \right).
	\label{eq:MeanJ}
\end{equation}
The amplitude of the turbulent resistivity $\beta$ and that of the $\gamma$ term related to the turbulent cross-helicity control the Lorentz force around the diffusion region of reconnection where they are finite.
For high Reynolds number plasmas, the turbulent Reynolds number $R_\beta\sim 1/ \beta$ is lower than the
molecular one $R_\eta\sim1/\eta$. In such a situation, the Lorentz force is decreased by an increased turbulence. The size of the diffusion region is enlarged and the reconnection rate is enhanced. The turbulent
heicity $H$ related to the $\alpha$ term does not enter
\eref{eq:MeanJ} directly but through its effect on the production of the turbulent energy and the turbulent cross-helicity which are both related to the $\beta$ and $\gamma$ terms. The turbulent helicity $H$ reduces
the strength of the turbulent energy $K$. The turbulent resistivity $\beta$ is reduced and the Lorentz force is increased. As a result, the size of the diffusion region is diminished and the reconnection rate is slowed down.
This way the mean Lorentz force, as well as \eref{eq:MaBg}, is directly related to the turbulence dynamics. \\
\indent Our simulations show that the turbulent energy $K$ is located near the mean current density $\boldsymbol{J}$ concentration. Reconnection is enhanced by the turbulent resistivity $\beta$ related
to $K$. The cross-helicity $W$, on the other hand, appears to be distributed around the current density maxima due to the mean vorticity $\boldsymbol{\Omega}$. This is true for all initial equilibria considered (\fref{fig:SpatialTurbuKW}) as theoretically predicted.\cite{Yokoi4} \\
\indent In addition to the energy and cross-helicity of the turbulence, a turbulent
helicity is generated as soon as an out-of-plane guide magnetic field is considered. According to its definition [\eref{eq:MathHKW2}], the total mean turbulent helicity consists of kinetic and magnetic contributions
\begin{eqnarray}
	H_{tot} &=& -H_{kin}+H_{mag}\label{eq:Htot}\\
            	&=& -\left(\Mean{V\cdot \Omega}-\Mean{V}\cdot\Mean{\Omega}\right) + \left(\frac{\Mean{B\cdot J}-\Mean{B}\cdot\Mean{J}}{\bar\rho}\right)
\end{eqnarray}
An anti-parallel Harris-type CS equilibrium does not produce any turbulent helicity due to mirror-symmetry. A guide magnetic field can be added, however, to a Harris-type CS without changing its equilibrium. It produces initially a turbulent
magnetic helicity $H_{mag}$ due to the alignment of the guide field and the mean current density (mirror-symmetry broken). The initial force-free CS equilibrium produces a force directed out of the reconnection plane which aligns of the mean velocity and vorticity field. It generates a kinetic helicity $H_{kin}$ in addition to a magnetic helicity $H_{mag}$.
The initial conditions for the force-free equilibrium produced, therefore, both kinetic ($H_{kin}$) and magnetic ($H_{mag}$) turbulent helicity while
a Harris-type CS equilibrium with guide field initially only generates a turbulent magnetic helicity ($H_{mag}$). Even though in a Harris-type CS with guide field there is no turbulent kinetic helicity present initially, it is later generated during the non-linear evolution of the reconnecting current sheet (\fref{fig:SpatialTurbuHel}). In both Harris-type and force-free CSs, the turbulent magnetic and kinetic helicity are located mainly at
and near the 'X'-points of reconnection. Its location at the 'X'-point is due to the magnetic contribution $H_{mag}$. On the other hand, the distribution of the total turbulent helicity near the 'O'-points is a consequence  of its kinetic contribution $H_{kin}$.
Hence, the guide magnetic field is the reason for an increase of the total turbulent helicity $H_{tot}$. This relates the maximum reconnection rate to the guide field strength. \\
\indent A strong guide field slows the reconnection rate [\eref{eq:MaBg}].
This reduction can be attributed to the turbulent helicity $H$ related to the $\alpha$ term in \eref{eq:EMFY}.
Its influence on the rate of magnetic reconnection can be obtained by the Alfv\'en Mach number $M_A$. Supposing steady state
reconnection at each 'X'-point, a dimensional analysis reveals
\begin{equation}
M_A^2=\eta_\ast+\beta_\ast\left(1-\frac{|\gamma_\ast|+|\alpha_\ast|}{\beta_\ast}\eta_\ast^{1/2}\right).
\label{eq:MachA}
\end{equation}
The $_\ast$ indicates that only the dimensions of the variables are used for the derivation.
The normalisation is given by the Alfv\'en speed $V_A$ and the half-width $L_0$ for $\eta_\ast$, $\beta_\ast$ and $\gamma_\ast$. The $\alpha_\ast$
term is normalised by $V_A$. The reconnection rate decreases as soon as the $\alpha_\ast$ term is finite. This effect can be traced back to
\eref{eq:EMFY}, where the influence of the turbulent resistivity ($\beta_\ast$) is attenuated by the
turbulent helicity ($\alpha_\ast$).
In fact, the turbulent helicity, as well as the turbulent cross-helicity, reduces the production of turbulent energy.\cite{Yokoi1} The cross-helicity localises the turbulent energy near the 'X'-points in the diffusion region by suppressing its production around it. It is further suppressed by the turbulent helicity at the 'X'-points. This suppression of the apparent turbulent resistivity $\beta$ reduces the reconnection rate. This relates the rate of energy conversion in guide field reconnection to the turbulence dynamics.
\begin{figure}[h]
	\centering
	\begin{tabular}{cc}
		\hspace{-0.5cm} {\includegraphics[width=0.5\linewidth,keepaspectratio]{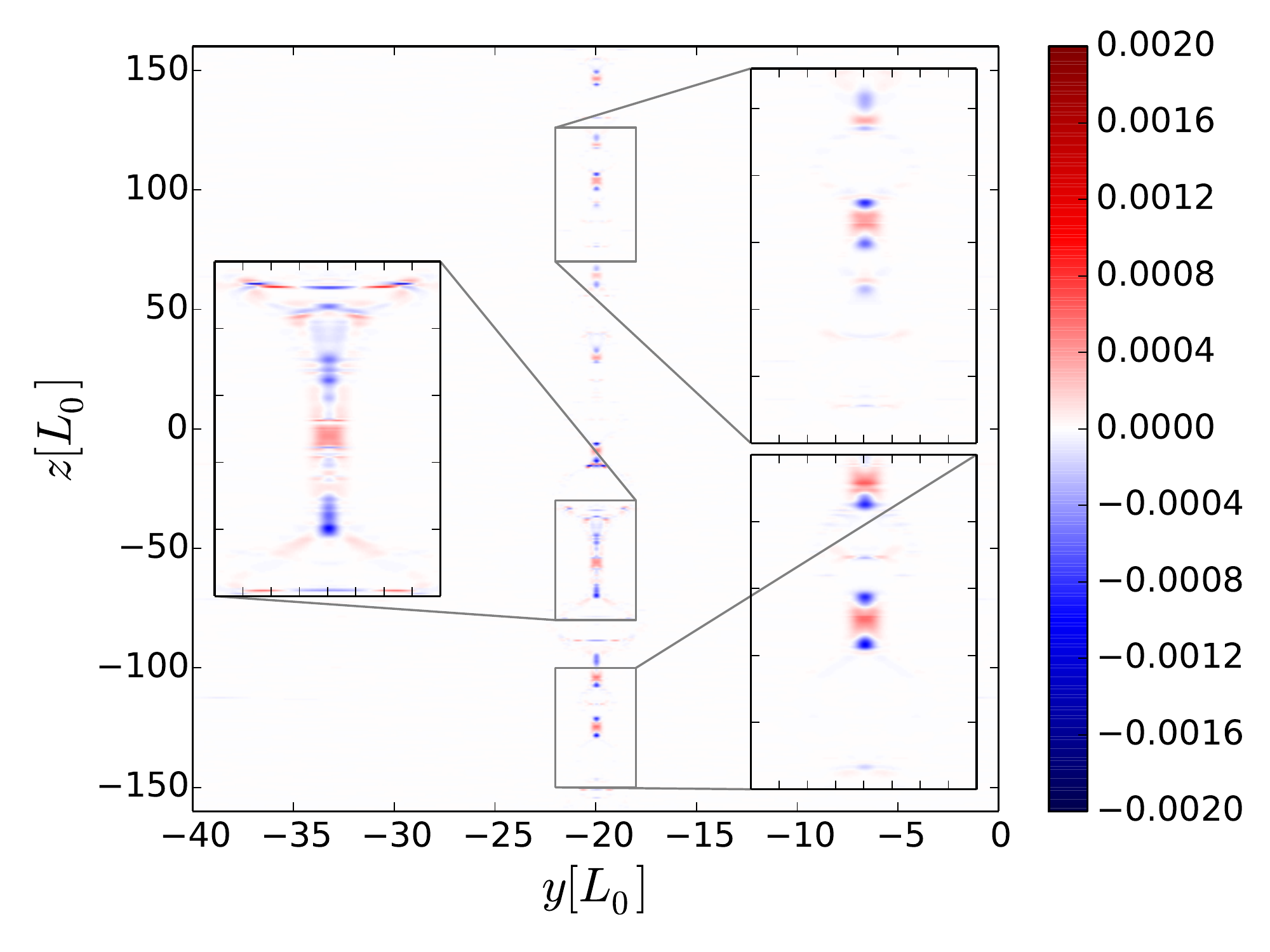}} &{\includegraphics[width=0.5\linewidth,keepaspectratio]{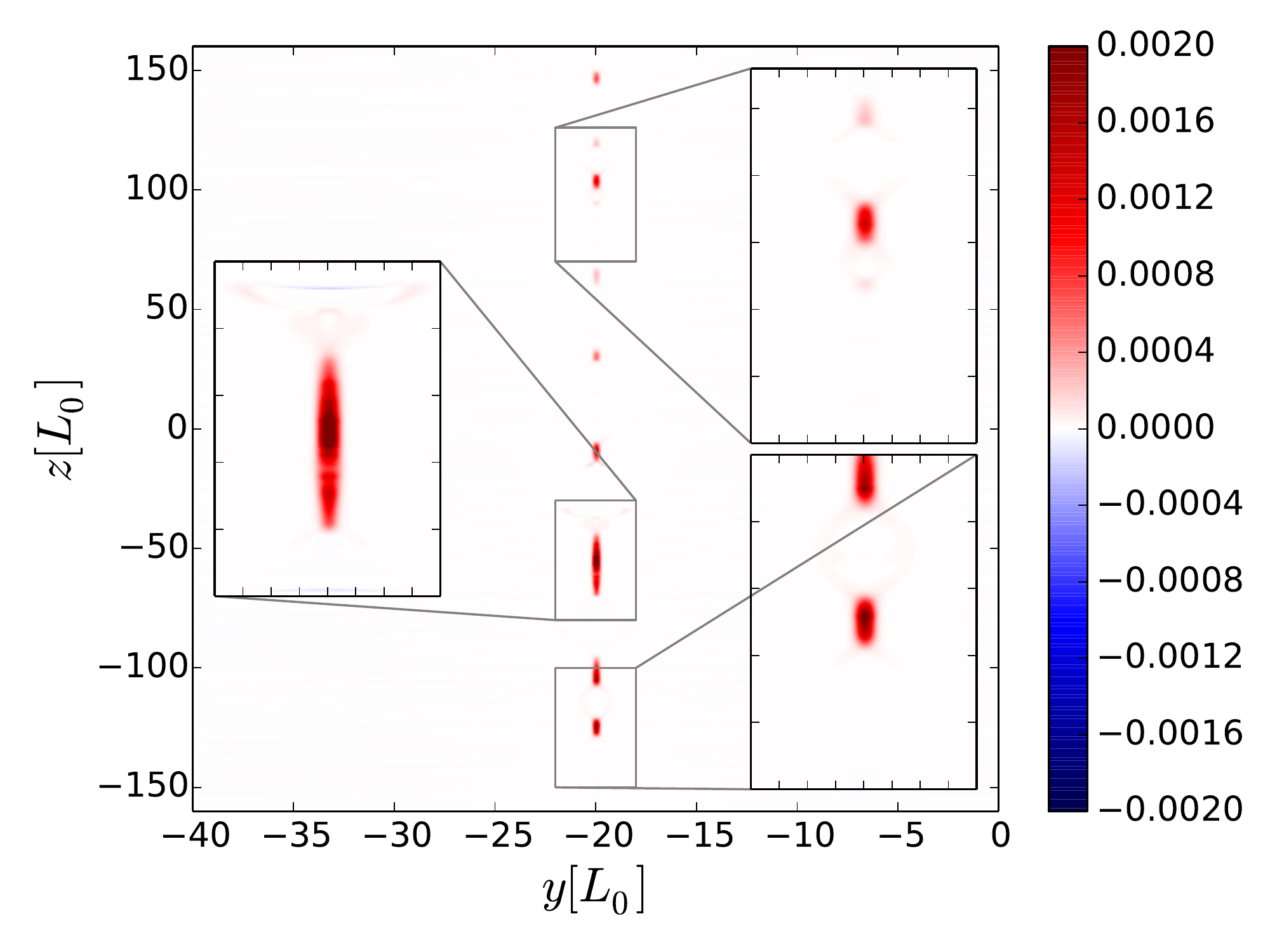}}\\
		\hspace{-0.5cm} a) ${\cal{E}}$, Harris $b_g=0$ & b) Model ${\cal{E}}_M$, Harris $b_g=0$\\
		\hspace{-0.5cm} {\includegraphics[width=0.5\linewidth,keepaspectratio]{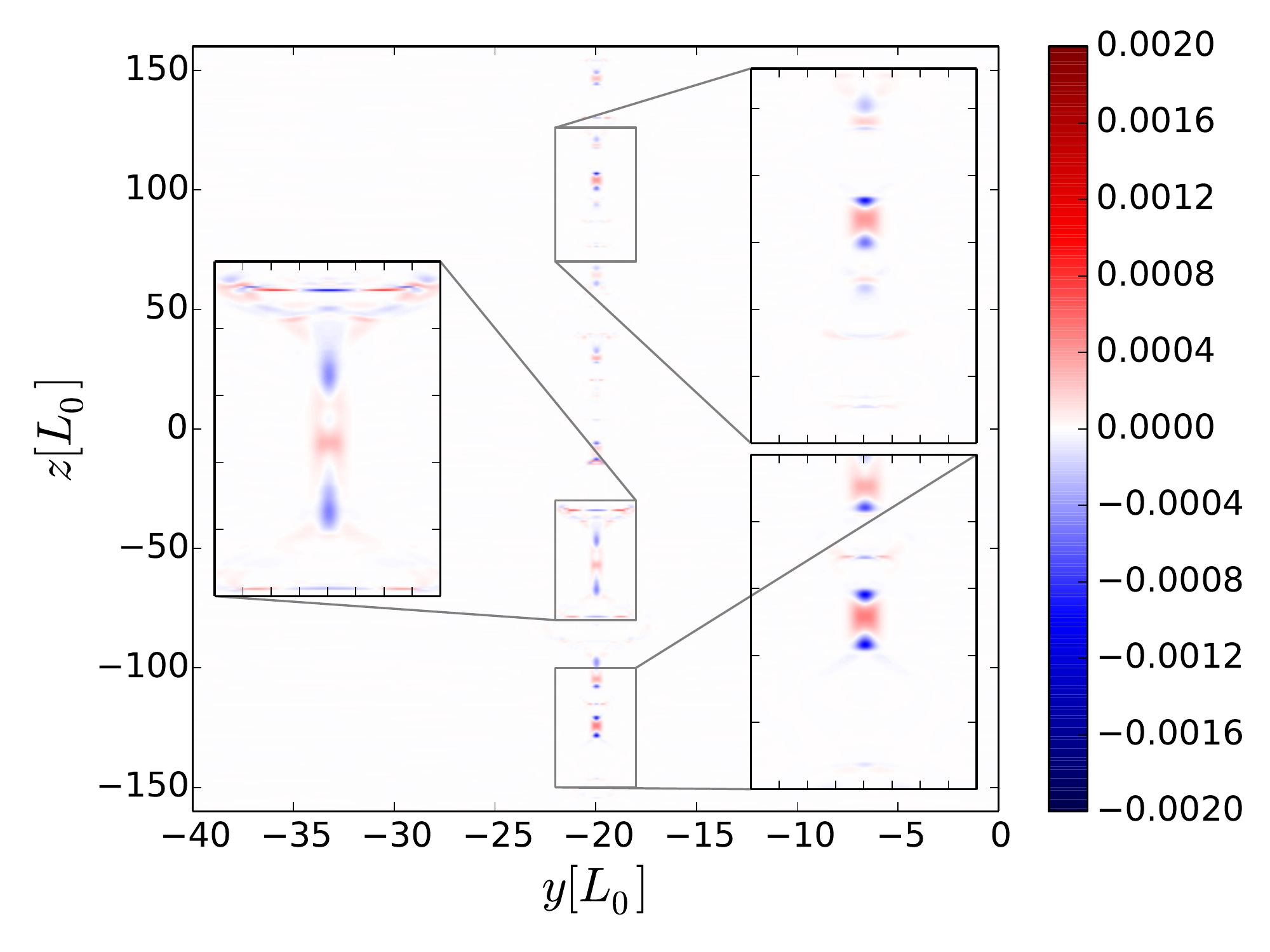}} & { \includegraphics[width=0.5\linewidth,keepaspectratio]{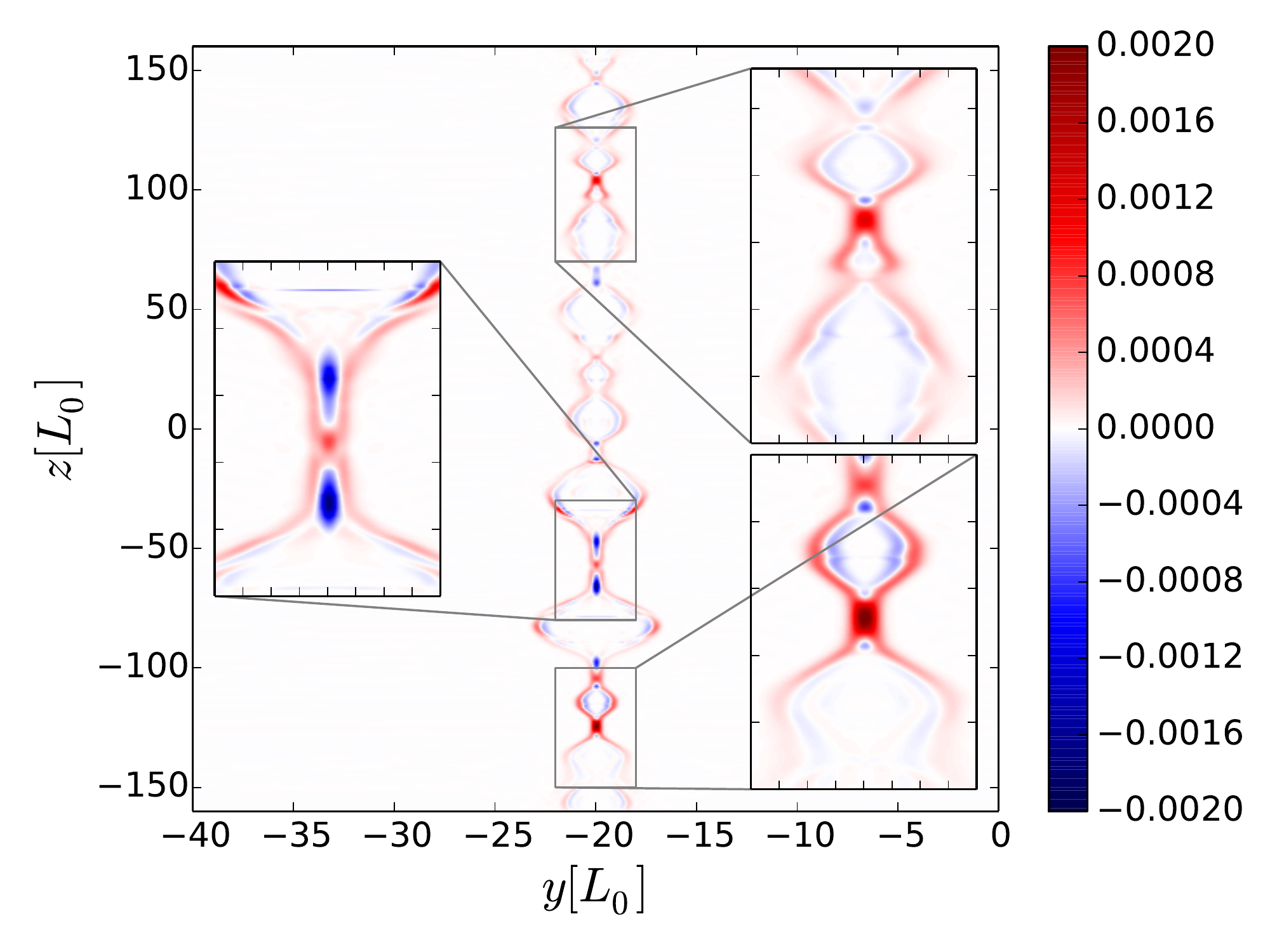}}\\
		\hspace{-0.5cm} c) ${\cal{E}}$, force-free $b_g=2$ & d) Model ${\cal{E}}_M$, force-free $b_g=2$\\
		\end{tabular}
		\caption{Spatial distribution at $100t/\tau_A$ of the electromotive force $\cal{E}$ and its modelling ${\cal{E}}_M$ in Harris-type CS $b_g=0$ a) and b)) and force-free CS $b_g=2$ ( c) and d)). The intensity of ${\cal{E}}$ is multiplied by a factor of two for visualisation purposes.}
\label{fig:EMF}
\end{figure} 
\subsection{Electromotive force}
\begin{figure}[h]
  \centering
  \begin{tabular}{cc}
	
	 \hspace{-0.5cm}  {\includegraphics[width=0.5\linewidth,keepaspectratio]{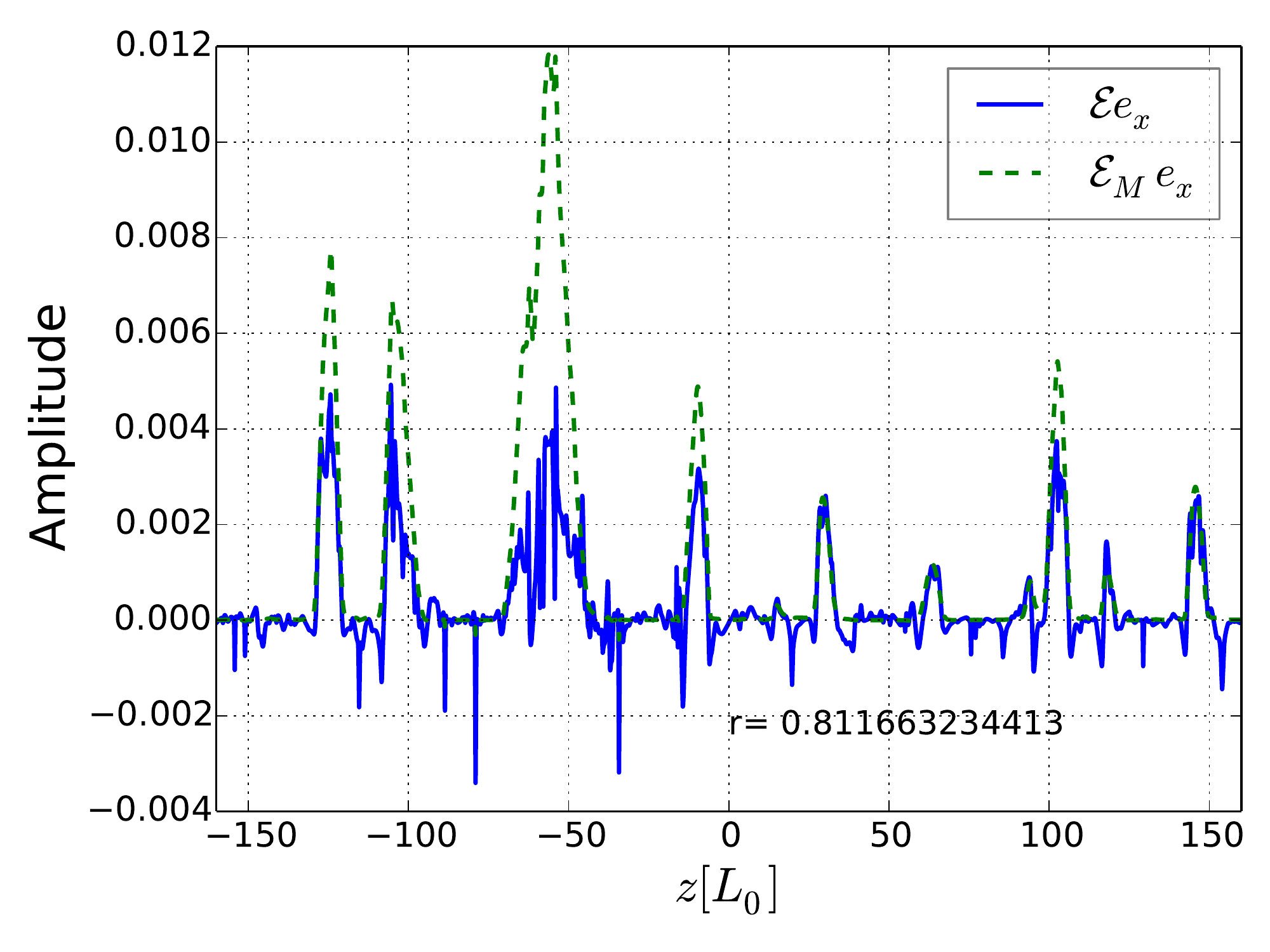}}&{\includegraphics[width=0.5\linewidth,keepaspectratio]{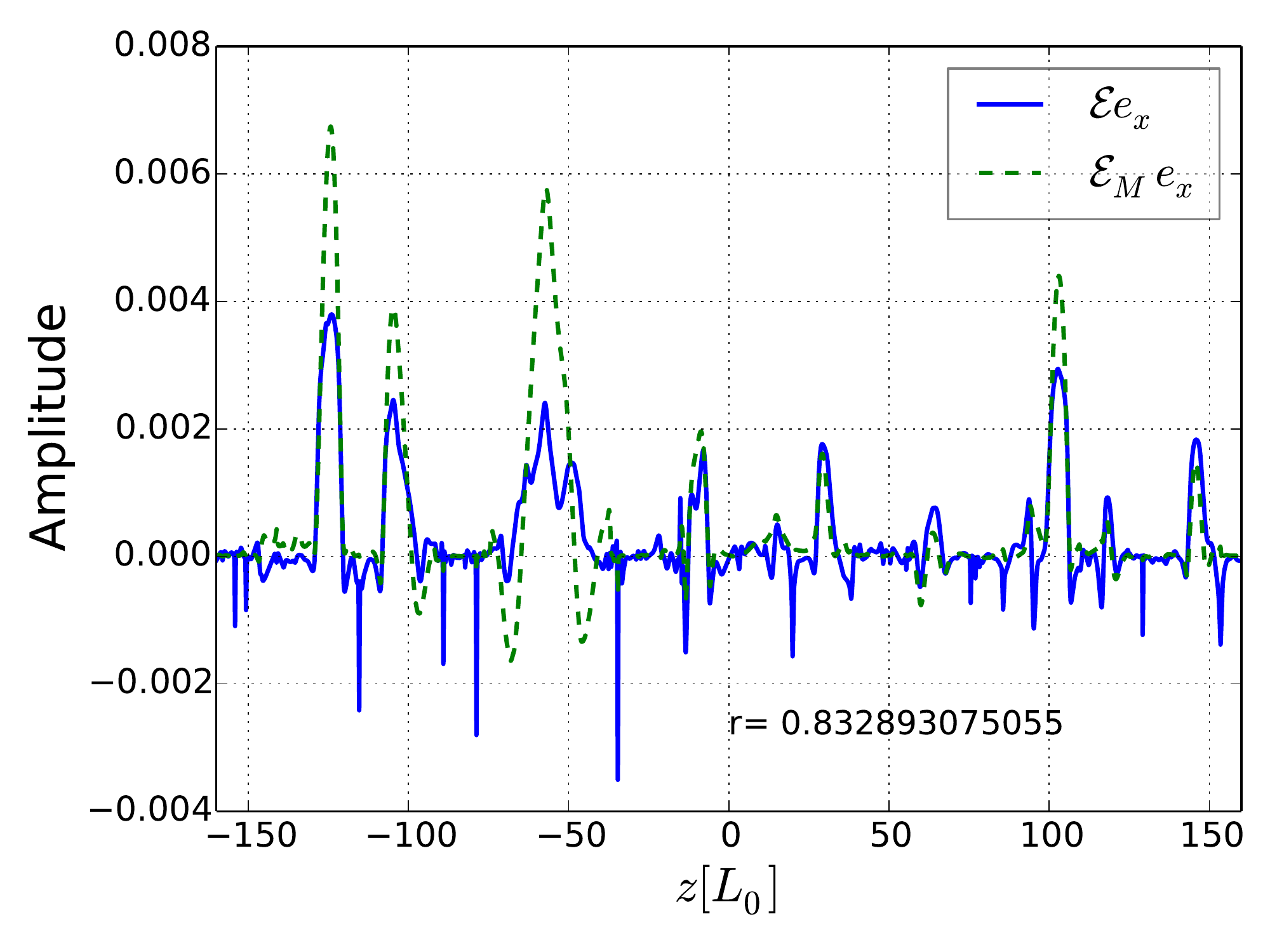}}\\
	 \hspace{-0.5cm}  a) Harris $b_g=0$, $r=0.81$ & b) Force-free $b_g=2$, $r=0.83$\\

  \end{tabular}	
  \caption{Comparison of the electromotive forces along the center of the CS at $t=100\tau_A$ with correlation factor $r$ for a) Harris-type CS $b_g=0$, b) force-free $b_g=2$. Solid line: ${\cal{E}}$. Dashed line: ${\cal{E}}_M$. The model ${\cal{E}}_M$ is rescaled by 3 to be in the range of the electromotive force definition ${\cal{E}}$.}
\label{fig:CompEMF}
\end{figure}

The electromotive force $\cal{E}$ is compared to the model ${\cal{E}}_M$ (\fref{fig:EMF}). In anti-parallel Harris-type
CS, the ${\cal{E}}_M$ is located at and around the 'X'-points similarly to the electromotive force ${\cal{E}}$. The amplitude of ${\cal{E}}_M$ is of the same order as the mean electromotive force but does, however, not reproduce the negative value near the
'O'-points. As soon as an out-of-plane guide magnetic field is considered in Harris-type CS, both the negative
and positive signs of the electromotive force are recovered by ${\cal{E}}_M$. The turbulent helicity, generated after the mirror-symmetry breakage by the guide magnetic field, contributes to the negative sign of ${\cal{E}}_M$. The model ${\cal{E}}_M$ is, however, overestimating the
electromotive force ${\cal{E}}$ calculation by a factor of three. The force-free CS shows similar results. Two reasons can be responsible for the overestimation. First the constants $C_\alpha$, $C_\beta$ and $C_\gamma$ influence the result. While $C_\beta$ and $C_\gamma$ are well established for the model under consideration, the value of $C_\alpha$ is not well known.\cite{1992PhFl....4..441H} On the other hand, the same turbulence timescale $\tau$ is chosen for all three turbulence variables $\beta$, $\gamma$ and $\alpha$.
In fact it should be defined by the turbulence dynamics itself.
On average over all reconnection site, the model electromotive force ${\cal{E}}_M$
corresponds to the behavior of $\cal{E}$ (\fref{fig:CompEMF}). 

\subsection{Energy Consideration}
\begin{figure}[h]
	\centering
	\begin{tabular}{cc}
	\hspace{-0.5cm}	{\includegraphics[width=0.5\linewidth,keepaspectratio]{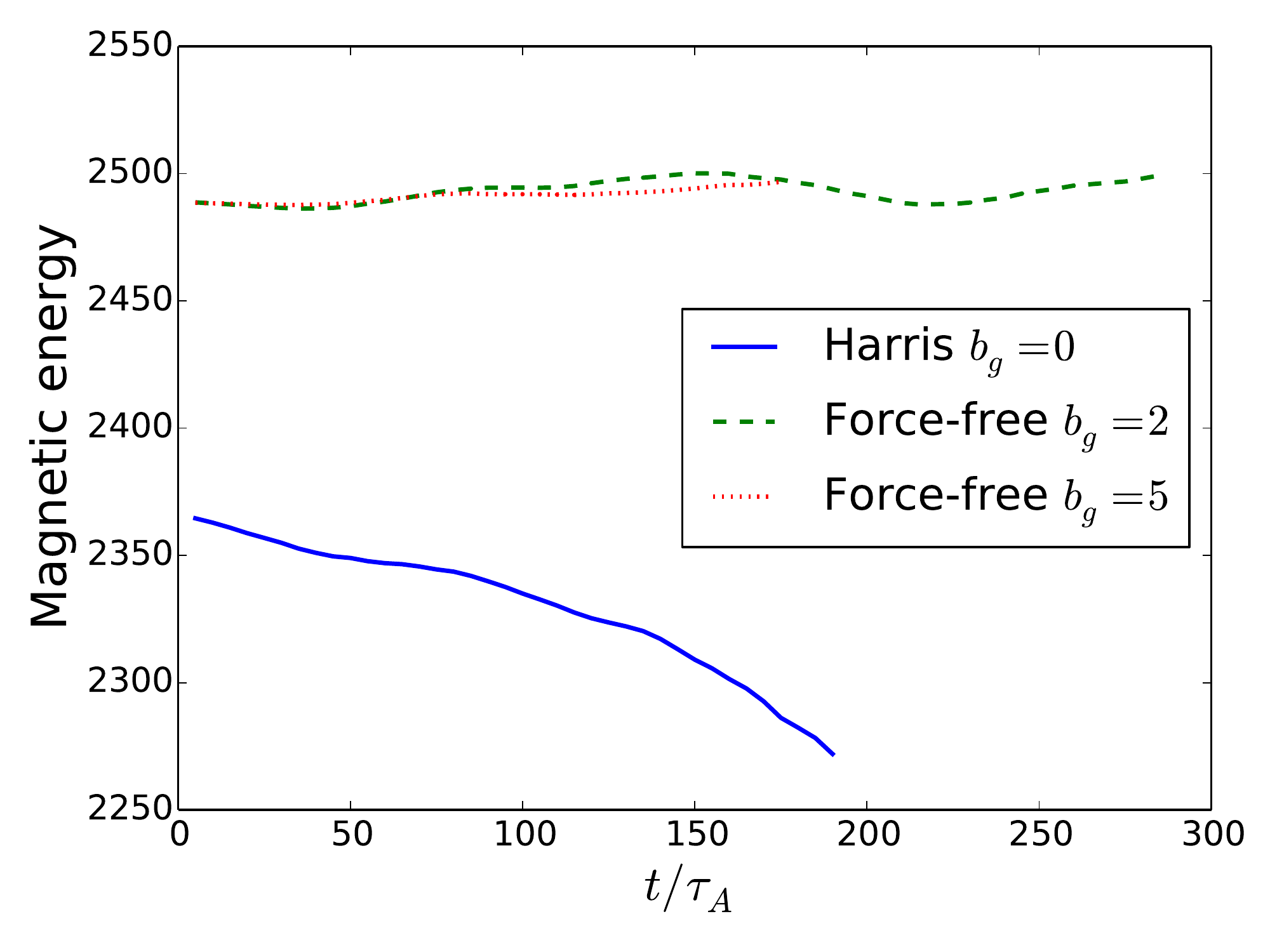}} & {\includegraphics[width=0.5\linewidth,keepaspectratio]{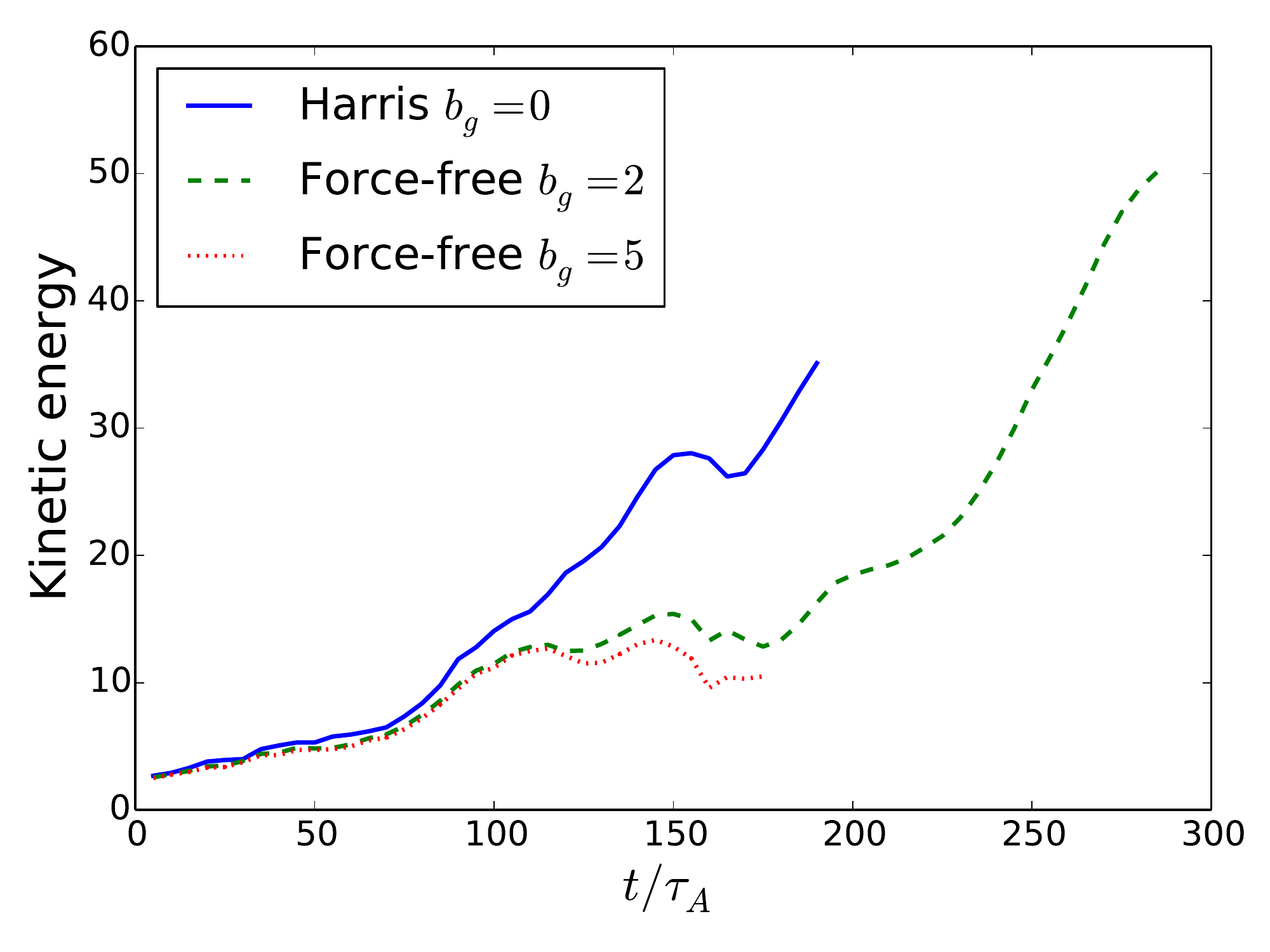}} \\
	\hspace{-0.5cm} a) Magnetic energy & b) Kinetic energy \\
	\end{tabular}
	\caption{Time history of a) the total magnetic energy and b) the total kinetic energy. The energies are computed for the Harris-type current sheet without guide field and force-free current sheet with guide magnetic field $b_g=2$ and 5.}
	\label{fig:EnergyTot}
\end{figure}

In our Gaussian filter approach, the mean energy density of the magnetic field $\overline{{\bf{B}}^2}/2\mu_0$ can be split into its mean $\overline{{\bf{B}}}^2$ and its fluctuation part $\overline{{\bf{b}}'^2}$. The evolution equation for the former is
\begin{equation}
	\frac{d}{dt}\int\limits_V \frac{\overline{\bf{B}}^2}{2\mu_0}d\boldsymbol{x}= \int\limits_V\left[-\eta\overline{\bf{J}}^2-\overline{\bf{V}}\cdot(\overline{\bf{J}}\times\overline{\bf{B}})+\overline{\bf{J}}\cdot{\cal{E}}\right]\ d\boldsymbol{x},
\label{eq:EnerMean}
\end{equation}
where, depending on its sign, the last term on the right-hand side may be a source or a sink of energy. A stretching of field lines increases the magnetic energy
while a contraction decreases it. Figure~\ref{fig:EnergyTot} a) depicts the evolution of the total magnetic energy
and b) of the total kinetic energy within the simulation
box. 
Note that the magnetic energy of the force-free equilibria is rescaled by a factor of three to be in the same range as the
Harris-type CSs energy. While the magnetic energy rapidly decreases in Harris-type CSs without guide magnetic field, a force-free equilibrium with guide magnetic field retains a high level of magnetic energy. On the other hand, the plasma kinetic energy is lower for force-free CSs compared with anti-parallel Harris-type CS ($b_g=0$). \\

\begin{figure}[h]
	\centering

		\hspace{-0.5cm} {\includegraphics[width=\linewidth,keepaspectratio]{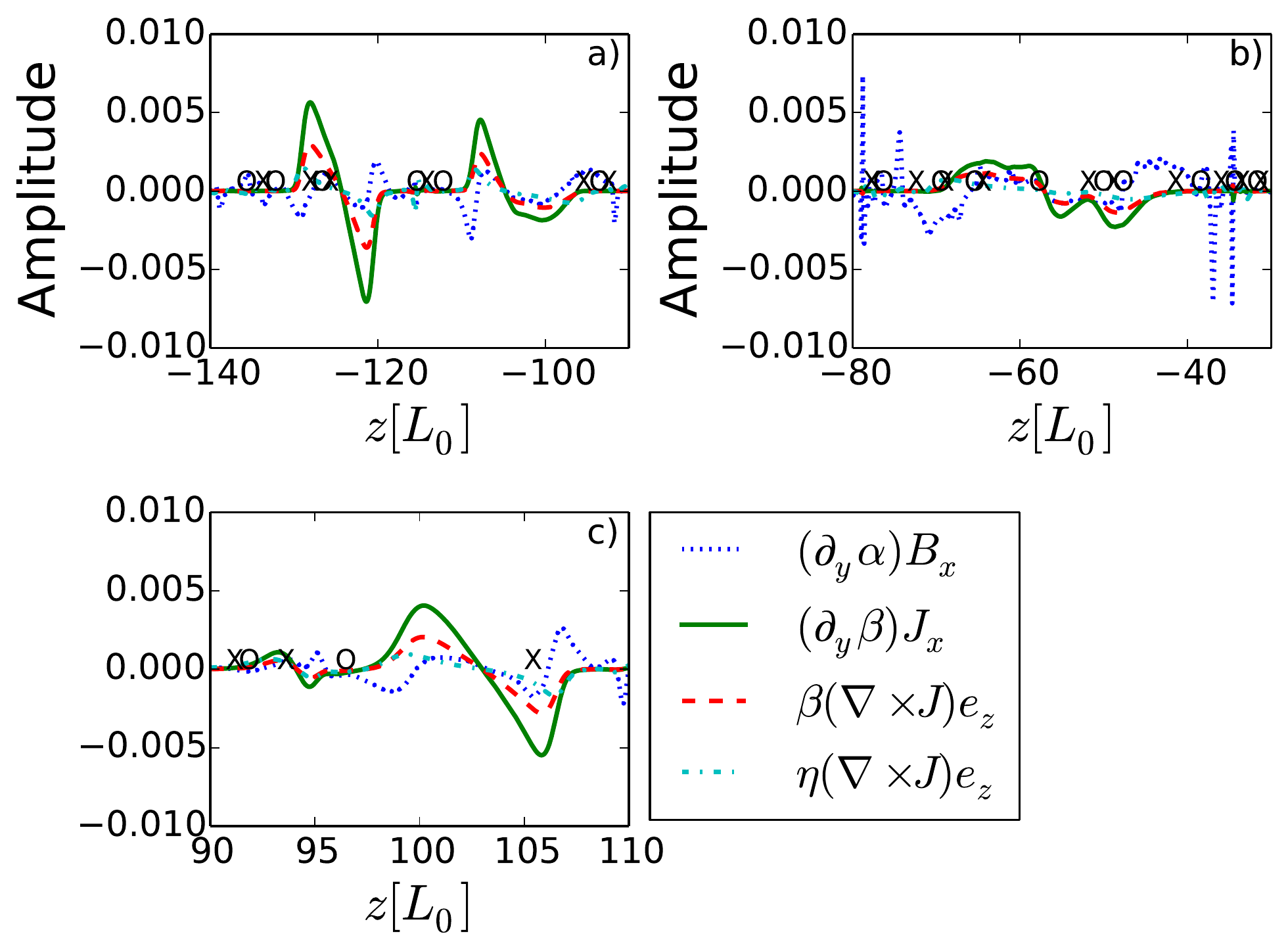}} \\
		\caption{Components of the diffusion and advection terms along the current sheet from the mean induction equation. They correspond to the zoomed regions of reconnection in \fref{fig:ContourFF2140}: a) the lower right zoom, b) the middle left zoom and c) the upper right zoom.  The \textbf{X} and \textbf{O} denote the 'X'- and 'O'-points. The amplitude is multiplied by 100.0 for visualisation purposes.}
	\label{fig:IndTermsFF2}
\end{figure}

The electromotive force is positive near 'X'-points and negative near 'O'-points (\fref{fig:EMF}). From turbulence viewpoint, the apparent turbulent resistivity $\beta$ is positive at and around the 'X'-points. Since
the mean current $\Mean{J}$ is negative in the present geometry, the first term in \eref{eq:EMFY} is positive at the 'X'-points where the current density accumulates the most. At the 'O'-point vicinity, the
residual turbulent helicity is found to be negative for a positive guide magnetic field, the last term in \eref{eq:EMFY} is then negative. The product $\overline{\bf{J}}\cdot\cal{E}$ is then negative close to the 'X'-points while it is positive close to the 'O'-points due to the balance of turbulence dynamics. The electromotive force causes a decrease of the magnetic energy at 'X'-points enhancing its conversion
into the kinetic energy and heat. Near the 'O'-points, it causes an increase of the magnetic energy, converting the kinetic (plasma flow) energy into the magnetic energy and heat. Hence, the total kinetic
energy is reduced and the total magnetic energy is increased there (\fref{fig:EnergyTot}).
The modeled ${\cal{E}}_M$ [\eref{eq:EMFY}] behaves similarly:
the turbulent energy related to the $\beta$ term enhances the annihilation of magnetic energy while the $\gamma$ term (depending on its sign) together with the $\alpha$ term acts like a source
term for the magnetic energy. The dynamical balance of turbulence modifies the contribution of the electromotive force to the mean magnetic energy. In \fref{fig:IndTermsFF2}, the different terms contributing to the
right-hand side of \eref{eq:IndYok} are shown for a given time as they are distributed along the
current sheet.
\Fref{fig:IndTermsFF2} presents the result for the force-free equilibrium with $b_g=2$ but similar results are
obtained for the other CS configurations and guide magnetic field strengths. The gradients of the turbulent helicity ($\alpha$ related term) and
the turbulent resistivity $\beta$ cause important effects. The $\alpha$ term acts against the turbulent ($\beta$) and molecular ($\eta$) resistivity.
In some locations, the turbulent helicity suppresses the turbulent diffusion ($\beta$ term), only the Joule
dissipation ($\eta$) can convert the magnetic energy into heat.
On the other hand, there is no mechanism to produce turbulent helicity in two dimensional Harris-type CSs without
guide field since mirror-symmetry is not broken. The turbulent cross-helicity is the only source term for the magnetic energy near 'O'-points.
The production of magnetic energy near the 'O'-points is, however, less than the counterpart in presence of turbulent helicity.
In such a situation, the annihilation of the mean magnetic field is enhanced by the turbulent resistivity $\beta$ because no
turbulent helicity ($\alpha$ related term), neither kinetic or magnetic, effects take place. There is no mechanism to reduce the
turbulent energy at the 'X'-points and magnetic reconnection can grow fast.

\subsection{Turbulence relation to mean fields}
\begin{figure}[h]
	\centering
	\begin{tabular}{cc}
	       
	        \hspace{-0.5cm}	{\includegraphics[width=0.5\linewidth,keepaspectratio]{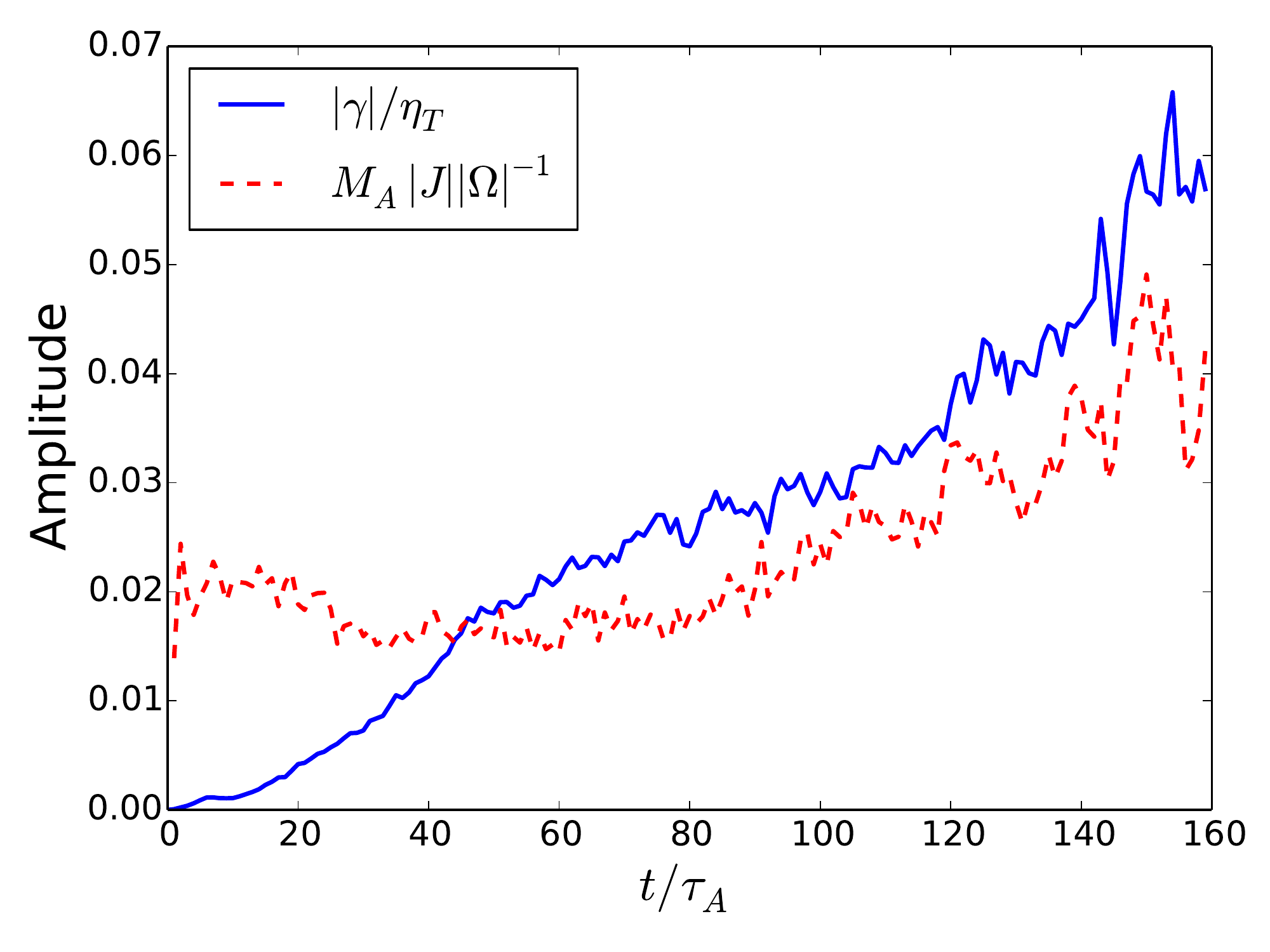}} & {\includegraphics[width=0.5\linewidth,keepaspectratio]{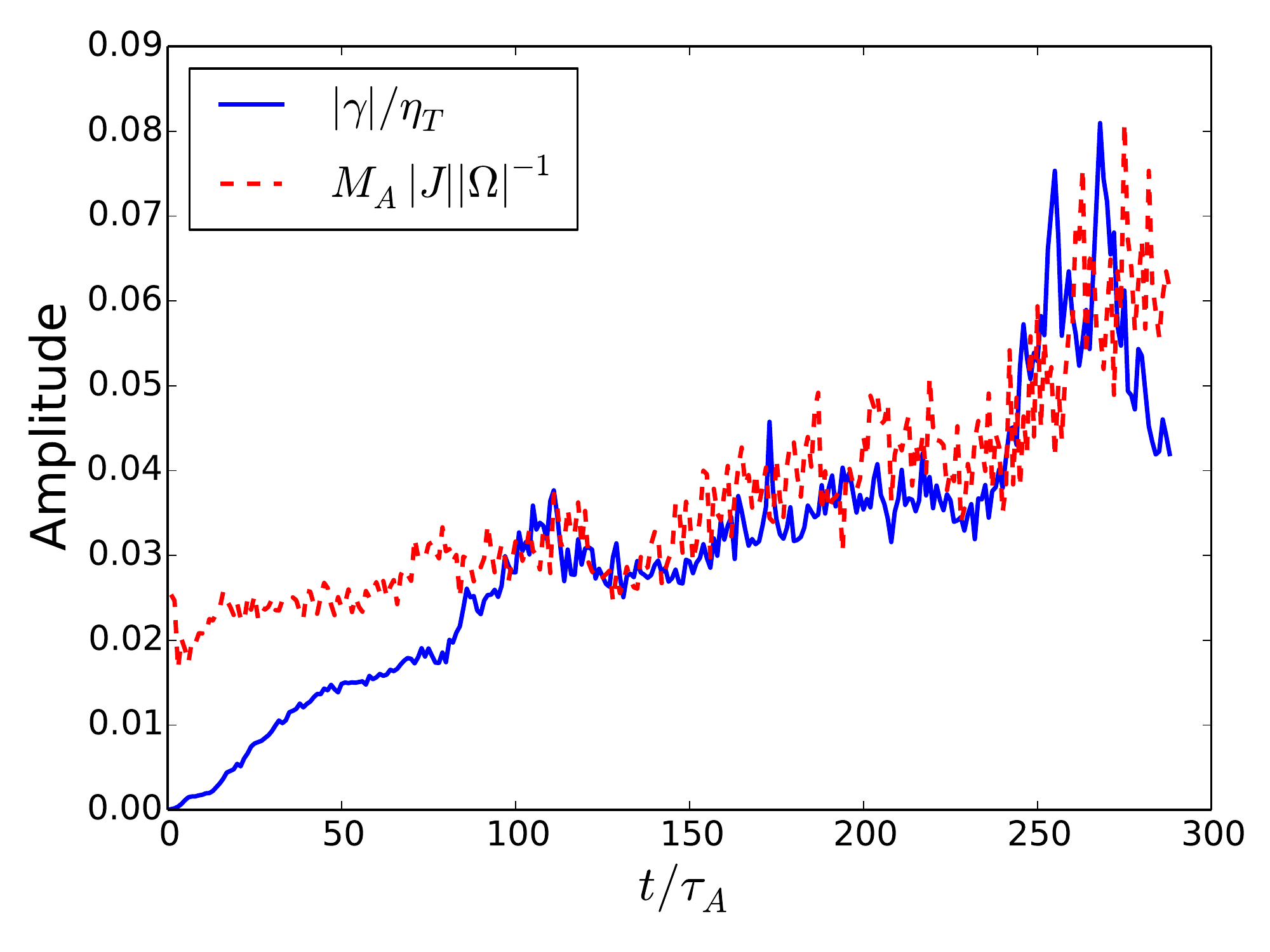}} \\
		\hspace{-0.5cm} a) & b)\\	
	\end{tabular}
	\caption{Time history of the amount of turbulence in the system according to the RANS model. a) Harris-type and b) Force free with $b_g=2$. Dashed line: calculated from \eref{eq:AmountTurbu}. Solid line: the ratio of the filtered turbulence quantities}
	\label{fig:AmountTurbu}
\end{figure}

\begin{figure}[h]
	\centering
	\begin{tabular}{cc}
		\hspace{-0.5cm} {\includegraphics[width=0.5\linewidth,keepaspectratio]{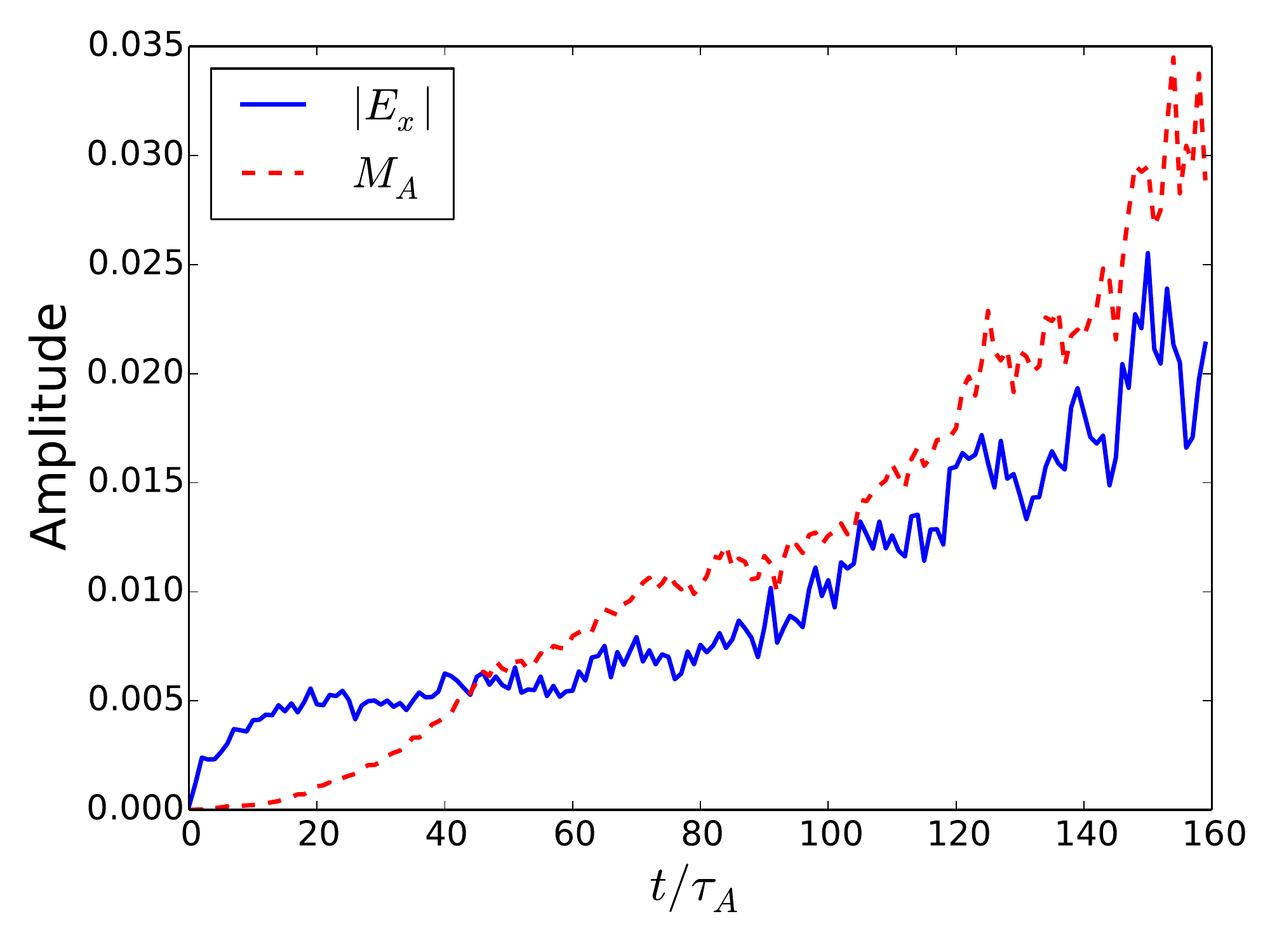}} &  {\includegraphics[width=0.5\linewidth,keepaspectratio]{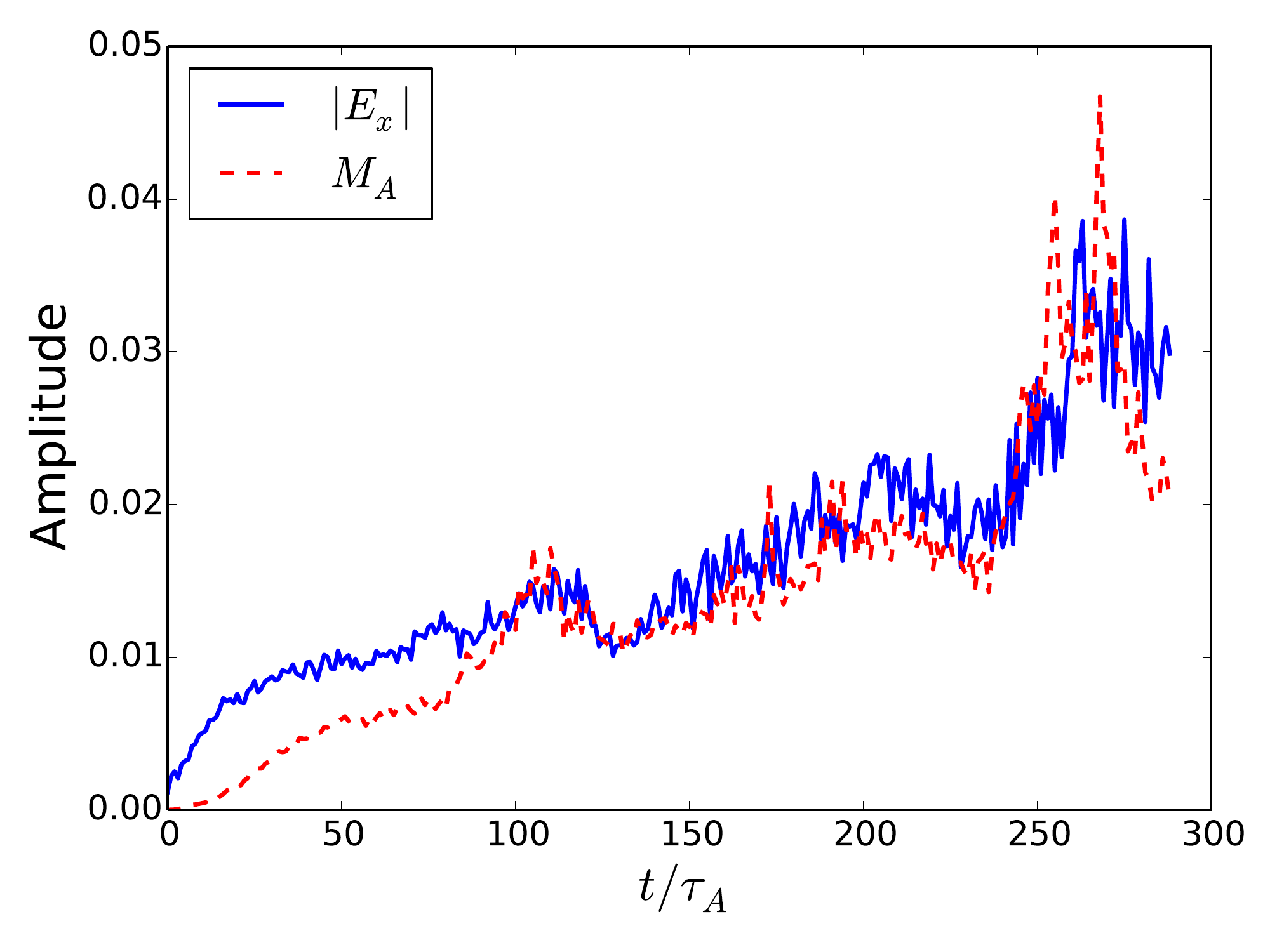}}\\
		\hspace{-0.5cm} a) Harris, $b_g=0$ & b) Force free, $b_g=2$ \\
	\end{tabular}
	\caption{Time history of the reconnection rate $|E_x|$ compared with the prediction $M_A$ of \eref{eq:MaTurb} for a) Harris-type current sheet with $b_g=2$ and b) force free equilibrium with $b_g=2$.}
	\label{fig:RecRatePred}
\end{figure}

\begin{figure}[h]
	\centering
		\hspace{-0.5cm} {\includegraphics[width=0.5\linewidth,keepaspectratio]{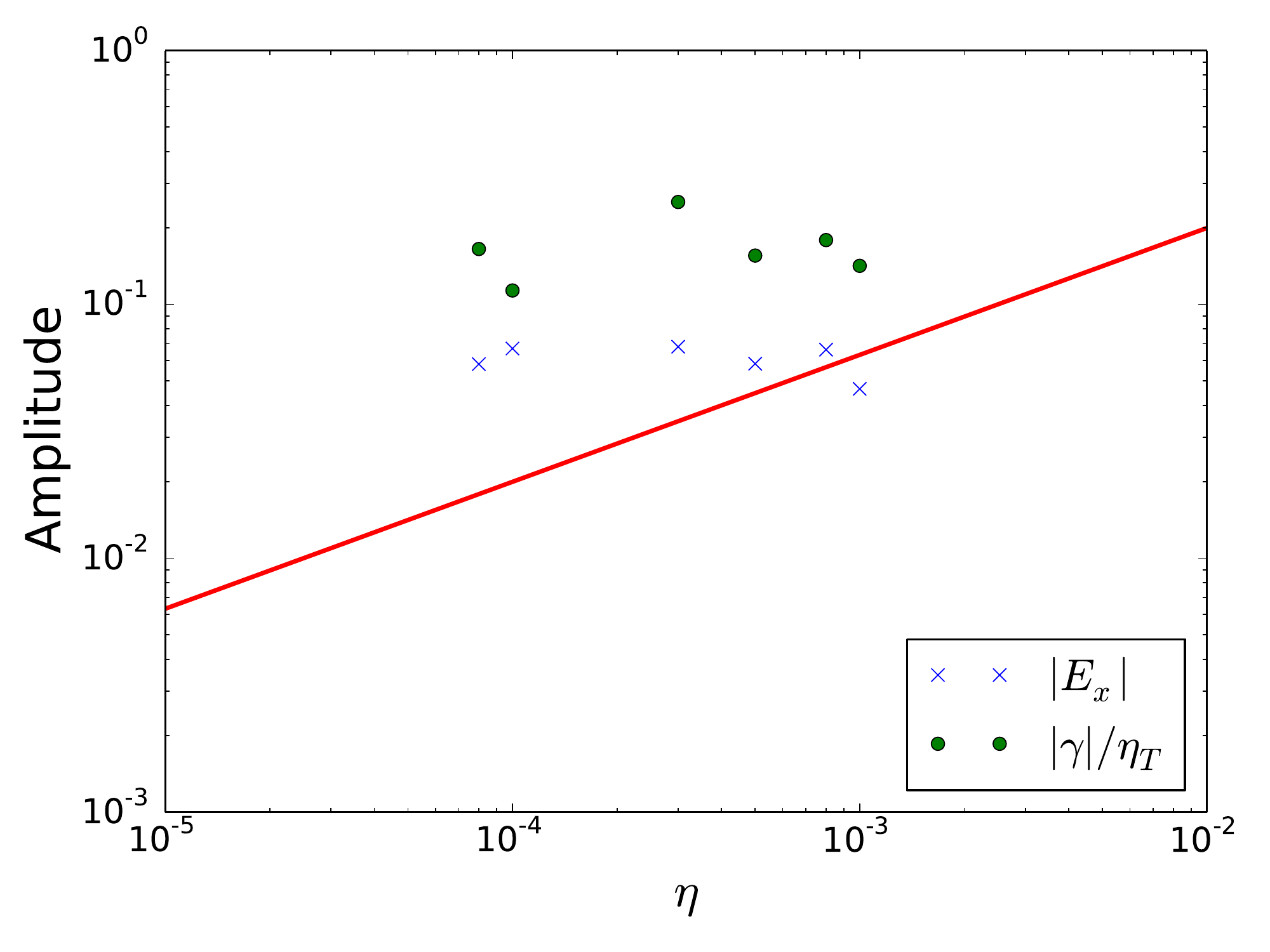}}
		\caption{Time average of the reconnection rate $|E_x|$ and amount of turbulence $|\gamma|/\eta_T$ [\eref{eq:AmountTurbu}] for different molecular resistivity $\eta$. There is a deviation from the Sweet-Parker scaling as $\eta$ decreases. The reconnection rate and the turbulence saturate similarly. Solid line: Sweet-Parker scaling.}
	\label{fig:RateDevSP}
\end{figure}

For a reconnecting current sheet, the relation between turbulence and the ratio $|\boldsymbol{\Omega}|/|\boldsymbol{J}|$ was shown to be related to the reconnection rate.\cite{Yokoi2,2015arXiv151104347W} The amount of turbulence in the system, represented by $|\gamma|/\beta$ in the Reynolds averaged
turbulence model, can be estimated as
\begin{equation}
	\left(\frac{|\gamma|}{\eta_T}\right) \cong M_A\frac{|\boldsymbol{J}_\star|}{|\boldsymbol{\Omega}|},
\label{eq:AmountTurbu}
\end{equation}
where $\boldsymbol{J}_\star=(\boldsymbol{J}\sqrt{\mu_0})/\sqrt{\rho}$ and $\eta_T=\eta +\beta$. In the limit of $\beta\gg\eta$, it is mainly the turbulent diffusivity $\beta$ that determines the denominator. In this limit, the
turbulence dominates the dissipation of magnetic energy in the diffusion regions [\eref{eq:MachA}].
The estimated amount of turbulence can be compared with the actual level obtained by the filtered $K$ and $W$. Figure~\ref{fig:AmountTurbu} shows the time history of the amount of turbulence estimated by \eref{eq:AmountTurbu} and calculated from the filtered $K$ and $W$ with $\tau=1$, $C_\beta=0.05$, $C_\gamma=0.04$ and $C_\alpha=0.02$.
For this set of parameters, the estimation (\ref{eq:AmountTurbu}) is in quiet good agreement with the ratio computed directly from $\gamma$ and $\beta$.
Finally, the reconnection rate determined as before
as the averaged out-of-plane electric field $|E_x|$ at the 'X'-points is compared with  
\begin{equation}
M_A=\frac{|\boldsymbol\Omega|}{|\boldsymbol{J}_\star|}\left(\frac{|\gamma|}{\eta_T}\right).
\label{eq:MaTurb}
\end{equation} 
The reconnection given by \eref{eq:MaTurb} corresponds well with the value
directly obtained by the out-of-plane electric field $|E_x|$ (\fref{fig:RecRatePred}).
According to the Sweet-Parker (SP) scaling, the reconnection rate reduces as $\eta$ decreases: $M_A\propto S^{-1/2}\sim \eta^{1/2}$. Long current sheets unstable to plasmoid instability show, however, an independence of the reconnection rate with respect to the Reynolds number.\cite{bhattacharjee2009fast} Since turbulence is ubiquitous at large Reynolds number plasmas (small $\eta$), the deviation from the SP scaling can be attributed to turbulence. Following \eref{eq:MachA}, the reconnection rate is determined by turbulence at small molecular resistivity $\eta$. \Fref{fig:RateDevSP} shows the deviation from the SP scaling (solid line) of the reconnection as well as the amount of turbulence calculated as \eref{eq:AmountTurbu}. The turbulence saturates and the deviation from the SP scaling can be attributed to turbulence as well.

\section{Discussion and conclusions}
\label{Conclusions}
\begin{figure}[h]
	\centering
		\hspace{-0.5cm} {\includegraphics[width=0.5\linewidth,keepaspectratio]{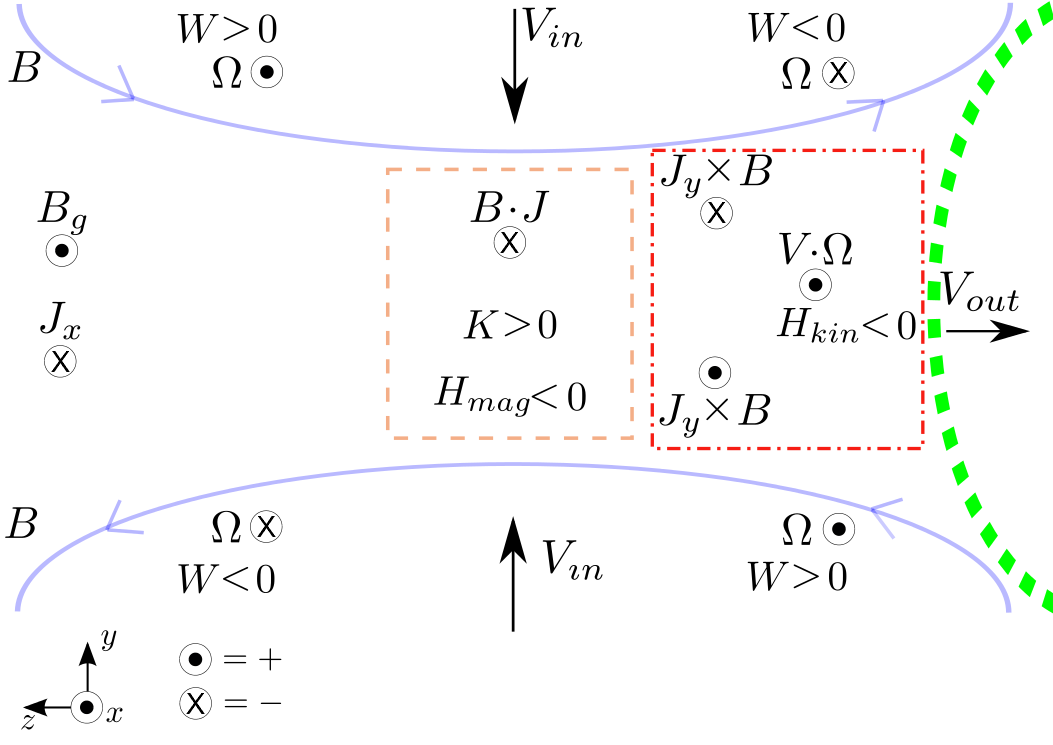}}
		\caption{Schematic representation of an 'X'-point. The turbulent cross-helicity $W$ has the same sign as the mean vorticity $\Omega$. The dashed box represents the region near the 'X'-point. There the turbulent magnetic helicity is negative as $\Mean{B}\cdot\Mean{J}<0$ and the turbulent energy $K>0$. The line dashed box represents the region close to an 'O'-point (curved dashed line). The mean current $\Mean{J_y}$ produced by the guide magnetic
		field generates a force $\Mean{J_y}\times\Mean{B}$. It generates a velocity in the ${\bf{e_x}}$ direction. The turbulent kinetic helicity is negative since $-\Mean{V}\cdot\Mean{\Omega}<0$. }
	\label{fig:SchemXO}
\end{figure}

We utilized a Reynolds averaged turbulence model in order to investigate the 
influence of small scale MHD turbulence on the plasmoid instability of long current 
current sheets in weakly dissipative plasmas.
For this sake we first validated the applicability of this turbulence model by filtering the data obtained by high resolution simulations of plasmoid-unstable 
Harris-type and force-free current sheets in the presence of finite guide fields with different strength. 
We found that the energy of the turbulence $K$ is growing mainly near 'X'-points 
in the dissipation region of magnetic reconnection.
There it causes additional, apparent "turbulent resistivity" (a $\beta$ -effect). The cross-helicty of the turbulence $W$ is growing around the 'X'-points where it forms in a quadrupolar structure with signs 
following the mean vorticity. This constrains the turbulent resistivity $\beta$ near the 'X'-points, enhancing the rate of magnetic reconnection. The turbulent helicity is growing, following the out-of-the-reconnection-plane (guide-) magnetic field, both near the 'X'- and 'O'-points.
While near the 'X'-points the produced turbulent helicity is mainly magnetic ($H_{mag}$), it is mainly a kinetic turbulent helicity $H_{kin}$
which is produced at the 'O'-points (\fref{fig:SchemXO}). 
As a result the turbulent kinetic helicity converts near 'O'-points plasma kinetic energy 
into magnetic energy and heat. This increases the total mean magnetic energy which then retains a high level even though reconnection takes
place.
Note that the turbulent-helicity related $\alpha$ term can reduce the turbulent and resistive annihilation of magnetic flux near 'X'-points if a guide field
breaks the mirror-symmetry of a pure Harris-type current sheet with anti-parallel magnetic fields only. 
At 'X'-points the apparent effective turbulent resistivity $\beta$ can become, therefore, balanced by turbulent helicity effects which slows down the conversion 
of magnetic into kinetic energy and reduces the reconnection rate compared to the case of anti-parallel Harris-type current sheets with vanishing guide fields 
($b_g=0$). A reduction of the reconnection rate in presence of large guide magnetic field 
can, therefore, be explained by means of turbulent helicity. \\
\indent
The modelled electromotive force ${\cal{E}}_M$ which depends on the energy of the turbulence $K$, the turbulent cross-helicity $W$ and the turbulent 
helicity $H$, agrees with the statistically determined $\cal{E}$.
It is, however, about three times larger than the calculated electromotive force.
In fact, in the course of plasmoid reconnection many differently sized 
islands are formed while in our model we choose the same constant turbulence timescale for all reconnection sites without taking into account their 
correlation. As it previously was found for single 'X'-point, turbulent reconnection becomes fast if the turbulence time scale is of the order of the Alfv\'en crossing 
time $\tau_A$. \cite{Yokoi3,2015arXiv151104347W} A choice of constant turbulence time scale is, therefore, a good approximation. The mean field turbulence model was, therefore, found to apply not only to
the problem of single 'X'-point but also for cascading plasmoid-type reconnection. \\
\indent According to the model, the turbulence is driven by the inhomogeneities of the large scale (mean-) fields, current density and vorticity. 
The mean fields and the ratio of turbulent energy to the turbulent cross-helicity determine the reconnection rate calculated as the out-of-plane mean electric field at the 'X'-points. The deviation of the reconnection rate from the Sweet-Parker scaling is found to be related to the saturation of the SGS
turbulence. The proposed Reynolds-averaged turbulence model is, therefore, able to reproduce the consequences of SGS effects for the reconnection rate of
turbulent plasmoid-unstable current-sheet reconnection as well as its deviation from the Sweet-Parker scaling. The SGS turbulence model reproduces the macroscopic
electromotive force and explain its dependencies. The model also allows to describe guide magnetic field effects controlling the turbulent helicity $H$
and its influence on the energy conversion rate.
\appendix
\section{Effects of the Filter Width\label{App:fitlerWidth}}
\begin{figure}[h]
	\centering
	\hspace{-0.5cm} {\includegraphics[width=0.5\linewidth,keepaspectratio]{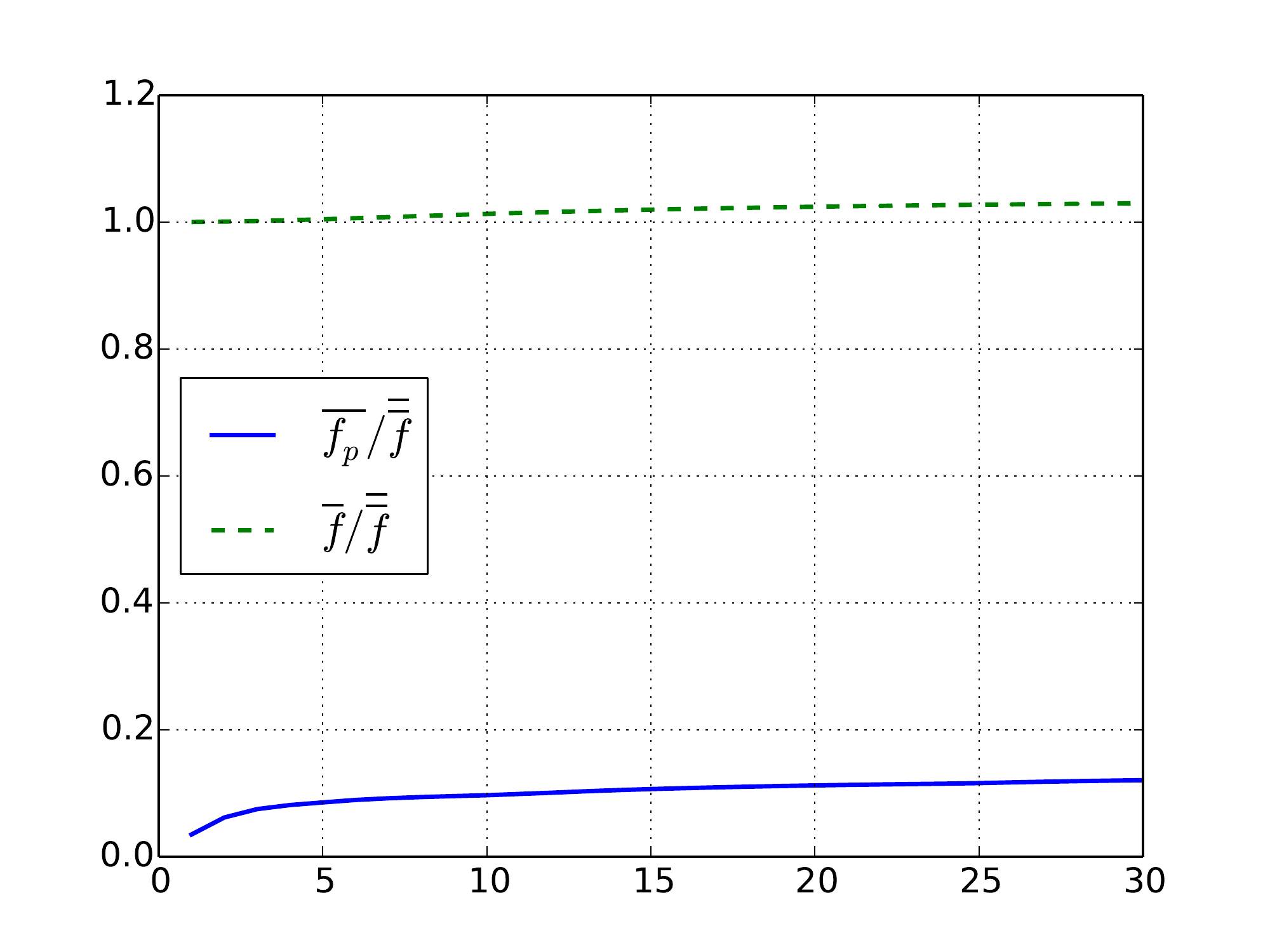}}
	\caption{Amplitude of the box average of the ratio $\overline{f'}/\overline{\overline{f}}$ and $\overline{f}/\overline{\overline{f}}$ with respect to the filter width. $f_p\equiv f'$}
	\label{fig:FiltWidt}
\end{figure}

The mean field quantities are calculated by means of a Gaussian filter. Such a filter cannot, however, strictly fulfill
Reynolds rules, i.e., $\left<f'\right>\neq 0$. This means that cross-terms such as $\Mean{v'\times B}$ may have 
important influence on the results. To avoid such issues, mean field are usually defined by global average.
This cannot be done in 2.5D simulations because the spatial variations of the mean fields describing the substructures of the current sheet
are required. A time average can not be carried out either, as a steady state is not reached properly. To reduce the effect of the
cross-terms, different filter widths have been tested. Applying the filter on the Reynolds decomposition
of a physical quantity $f$ gives
\begin{equation}
	\overline{f} = \overline{\overline{f}} + \overline{f'},
\end{equation}
where the over line correspond to a filtered quantity. The mean field is considered to be $\overline{f}$. To apply Reynolds rules,
$\overline{f'}/\overline{\overline{f}}$ should be as close to zero as possible and $\overline{f}/\overline{\overline{f}}$ should be close to one.
Figure~{\ref{fig:FiltWidt}} shows that an increased filter width increases both ratios values. In our calculation, the filter width was chosen such that these
ratios are close enough to fulfill Reynolds rules while turbulence quantities is still sufficiently spatially resolved. The chosen width in normalised unit is 5.

\section{Heuristic derivation of the Lorentz Force\label{App:LorForce}}
The electric field vanishes identically at the current sheet (CS) boundary. This provides $\overline{\bf{J}}\cong \overline{\bf{V}}\times\overline{\bf{B}}/\eta$. The Lorentz force components
across the CS for both guide field and non guide field equilibrium are (over lines are omitted)
\begin{equation}
			\overline{F_L}\bf{e_y}= \left\{ \begin{array}{rcl}
			\left(-V_yB_z^2+V_zB_yB_z-V_yB_x^2\right)/\eta & \mbox{for} & B_x \neq0 \\
			\left(-V_yB_z^2+V_zB_yB_z\right)/\eta & \mbox{for} & B_x =0 
					\end{array}\right.
					\label{eq:FLcompo}
\end{equation}
where the guide field is represented by the component $B_x$. When a magnetic field line changes its topology at the
'X'-point, it is assumed that $\overline{F_L}\bf{e_y}\equiv 0$. This condition yields
\begin{equation}
		\overline{F_L}\bf{e_y}= \left\{ \begin{array}{rcl}
			\frac{V_y}{V_z}=\frac{B_yB_z}{B_z^2+B_x^2} & \mbox{for} & B_x \neq0 \\
			\frac{V_y}{V_z}=\frac{B_y}{B_z} & \mbox{for} & B_x =0 
					\end{array}\right.
					\label{eq:FLcompoVanish}
\end{equation}
We finally obtain from the definition of the Alfv\'en Mach number $M_A=V_y^2/V_z^2$ the relation
\begin{equation}
	M_{A,b_g} = M_A\left(\frac{B_z^2}{B_z^2+B_x^2}\right)^2, 
	\label{eq:RecRatesBgs}
\end{equation}
where $M_{A,b_g}$ and $M_A$ are the reconnection rates for guide field and non-guide field CS equilibria. 
\begin{acknowledgments}
	One of the author (FW) acknowledges the International Max Planck Research School (IMPRS) at the University of G\"ottingen as well as the CRC 963 project A15. JB thanks the
	Max-Planck-Princeton Center for Plasma Physics for its support. 

\end{acknowledgments}

\bibliography{Arxiv}

\end{document}